\documentclass[lineno]{jfm}
\usepackage{graphicx}
\usepackage{placeins}
\usepackage{breqn}
\usepackage{caption}
\usepackage{subcaption}
\usepackage[usenames,dvipsnames]{xcolor}
\usepackage[colorinlistoftodos]{todonotes}
\usepackage{comment}
\usepackage{mathrsfs}
\usepackage{empheq}
\usepackage{csquotes}
\usepackage{siunitx}
\usepackage{arydshln}

\usepackage[dvipsnames]{xcolor}
\newcommand\myshade{85}
\colorlet{mylinkcolor}{YellowOrange}
\colorlet{mycitecolor}{MidnightBlue}
\colorlet{myurlcolor}{violet}
\usepackage[colorlinks,citecolor=blue]{hyperref}
\hypersetup{
  linkcolor  = mylinkcolor!\myshade!black,
  citecolor  = mycitecolor!\myshade!black,
  urlcolor   = myurlcolor!\myshade!black,
  colorlinks = true,
}
\newcommand{\blue}[1]{{\color{blue} #1}}

\shorttitle{Structures driving trailing-edge noise - Experimental investigation}
\shortauthor{S. Demange and others}

\title{Identification of structures driving trailing-edge noise. Part I - Experimental investigation}

\author{
Demange, S.\aff{1}, 
Yuan, Z.\aff{2},
Jekosch, S.\aff{3},
Sarradj, E.\aff{3},
Hanifi, A. V. G.\aff{2},
Cavalieri A.\aff{4}
\and
Oberleithner, K.\aff{1}
    \corresp{\email{s.demange@tu-berlin.de}}
}

\affiliation{
  \aff{1}Laboratory for Flow Instability and Dynamics, Technische Universität Berlin, 10623 Berlin, Germany
  \aff{2}FLOW, Department of Engineering Mechanics, KTH Royal Institute of Technology, Stockholm, Sweden
  \aff{3}Institute of Fluid Mechanics and Engineering Acoustics, Technische Universität Berlin, 10587 Berlin, Germany
  \aff{4}Divisao de Engenharia Aeronáutica, Instituto Tecnológico de Aeronáutica, São José dos Campos, Brazil
}

\begin{document}

\maketitle




\begin{abstract}{
Trailing-edge (TE) noise is the main contributor to the  acoustic signature of flows over airfoils. It originates from the interaction of turbulent structures in the airfoil boundary layer with the TE. This study experimentally identifies the flow structures responsible for TE noise by decomposing the data into spanwise modes and examining the impact of spanwise coherent structures on sound emission. We analyse a NACA0012 airfoil at moderate Reynolds numbers, ensuring broadband TE noise, and use synchronous measurements of surface and far-field acoustic pressure fluctuations with custom spanwise microphone arrays. Our results demonstrate the key role of coherent structures with large spanwise wavelengths in generating broadband TE noise. Spanwise modal decomposition of the acoustic field shows that only waves with spanwise wavenumbers below the acoustic wavenumber contribute to the radiated acoustic spectrum, consistent with theoretical scattering conditions. Moreover, a strong correlation is found between spanwise-coherent (zero wavenumber) flow structures and radiated acoustics. At frequencies corresponding to peak TE noise emission, the turbulent structures responsible for radiation exhibit strikingly large spanwise wavelengths, exceeding $60\%$ of the airfoil chord length. These findings have implications for numerical and experimental TE noise analysis and flow control. The correlation between spectrally decomposed turbulent fluctuations and TE noise paves the way for future aeroacoustic modelling through linearized mean field analysis. A companion paper further explores the nature of the spanwise-coherent structures using high-resolution numerical simulations of the same setup.
}
\end{abstract}
\section{Introduction}

Wind turbines are widely used as renewable power plants, providing a cheap, clean and sustainable source of energy. However, the acoustic footprint of wind turbines in operations strongly hinders the development of new onshore wind energy projects and compels many existing wind farms to operate at suboptimal conditions to limit acoustic nuisance. Among the various noise sources from wind turbines (e.g., generator noise, tip vortex noise, leading-edge turbulent flow noise), trailing-edge (TE) noise has been identified as the primary contributor of the acoustic spectrum within the audible frequency range~\citep{oerlemans_wind_2011}.

trailing-edge noise is generated by the scattering of surface pressure fluctuations (SPFs) that are caused by turbulent structures in the airfoil's boundary-layer~\citep{williamsAerodynamicSoundGeneration1970,brooksAirfoilSelfnoisePrediction1989}. In the case of low-to-moderate Reynolds numbers, the boundary layer is laminar or transitional, and TE noise manifests as a tonal phenomenon resulting from an acoustic-hydrodynamic feedback at discrete frequencies~\citep{Ricciardi2022}. Conversely, when the boundary layer over the airfoil is turbulent, either due to high Reynolds numbers or by forcing transition with tripping, trailing-edge noise appears as a broadband hump in the acoustic spectrum. The latter, which is sometimes referred to as turbulent boundary layer trailing-edge (TBL-TE) noise, is the subject of this study. The negative impacts of TBL-TE noise are not exclusive to wind turbines. It is also a significant issue in industries such as aviation, turbo-machinery, and the emerging field of drone propulsion~\citep{Arcondoulis2010}. Consequently, TE noise modelling remains an active area of research today. 

A recent review of the current state of the art on TE noise modelling is given by~\cite{lee_turbulent_2021}. The majority of models are derived from Lighthill's acoustic analogy~\citep{lighthill_sound_1952}, which describes the evolution of pressure in a fluid under the action of a source term known as Lighthill's stress tensor. These models often result in transfer functions that relate the power spectral density (PSD) of the surface pressure fluctuations (input) to that of the far-field acoustics (output)~\citep{williamsAerodynamicSoundGeneration1970}. A remaining challenge in TE noise modelling to date is to accurately predict the SPFs from the local turbulent flow without resorting to numerically expensive, highly detailed simulations. This becomes particularly complex when designing noise reduction techniques that interact with the turbulent flow~\citep{lee_turbulent_2021}. To better understand these interactions, as well as the underlying mechanism of TBL-TE noise, it is therefore useful to identify the turbulent structures that contribute to the acoustics, starting by characterizing their length scales.

When investigating the mechanism of TE noise, analyses typically consider two-dimensional airfoils free of end effects and assume the turbulent flow to be statistically independent of the spanwise position. The sound field radiated by a spanwise section of the geometry is then recovered from an integration of the source (input) term along the span. In the frequency domain, this leads to the explicit appearance of the spanwise coherence length, $\Lambda_z$, of the SPF spectrum in many of the TE noise models~\citep{amiet_acoustic_1975,howe_review_1978}. This parameter represents a characteristic statistical measure for the most energetic (or integral) spanwise length scale of the SPFs contents at a given frequency. Its relation to the physical length scales of flow structures generating TE noise will be discussed later in this paper.

Naturally, measurements of the coherence length $\Lambda_z$ have been included in experimental investigations of the relationship between SPFs and TE noise~\citep{brooks_trailing_1981, moreauEffectAirfoilAerodynamic2005,rozenbergRotatingBladeTrailingEdge2010,herrig_broadband_2013}. For instance, the study by~\cite{herrig_broadband_2013} provides comprehensive analyses of SPF spectra in the turbulent boundary layers over a NACA0012 airfoil. Their findings show that near the trailing edge, the coherence length is roughly similar to the boundary layer thickness, $\delta_{99}$, for low-to-intermediate frequencies, corresponding to chord-based Strouhal numbers in the range $St\in[1,\,15]$. 

Numerical studies such as~\cite{wang_computation_2000} have shown that the spanwise width of a simulated domain must be at least as large as $\Lambda_z$ to ensure statistically independent acoustic source regions. Additionally, large eddy simulations (LES) by~\cite{Wolf2012} demonstrated that spanwise coherence decays rapidly with increasing separation for high frequencies. Thus, the spanwise coherence length is a key input in many TE noise models and is frequently used to characterise the spanwise length scale of SPFs generating TE noise.

An alternative approach is to consider SPFs as a superposition of spanwise Fourier modes, analogous to the concept of incident sinusoidal gusts of~\cite{amiet_acoustic_1975}. In this context, remaining in the wavenumber domain provides a different perspective on the relevant spanwise length scales for TE noise. Assuming Fourier modes as the source term in Lighthill's acoustic analogy, \cite{Nogueira2017} derived an acoustic scattering condition for the wavenumber of the SPFs. It was established that only sources with spanwise wavenumbers less than or equal to the sonic wavenumber, are capable of radiating TE noise. Furthermore, their analysis highlights that SPFs with zero spanwise wavenumber should be significant for the acoustic radiation, given that they are propagative for all frequencies. This approach offers the potential for more specific modelling of TE noise by focusing on a limited set of acoustically relevant coherent structures, instead of considering the full turbulent spectrum. 

Building on the scattering condition, recent numerical analyses~\citep{Sano2019,Abreu2021} based on LES datasets of the turbulent flow around NACA0012 airfoils~\citep{Wolf2012,tanarro_effect_2020} have identified coherent structures linked to TBL-TE noise. These structures appeared as streamwise-elongated wavepackets in the boundary-layer. These analyses were limited to zero spanwise wavenumber, with the role of higher spanwise wavenumbers remaining unexplored. While non-zero (but small) spanwise wavenumbers are likely relevant to TE noise, resolving them  requires costly simulations with sufficiently wide spanwise domains. This specific aspects is explored in the companion paper of the present work~\citep{Yuan2024arxiv}.

A further benefit of a spanwise spectral analysis is its similarity with the ansatz from linear flow analyses. In particular, the resolvent framework~\citep{hwangLinearNonnormalEnergy2010,mckeonCriticallayerFrameworkTurbulent2010a} seems particularly well suited to TBL-TE noise, as it allows identifying coherent structures in turbulent flows associated with linear amplification mechanisms. Despite this potential, applications of the resolvent to the TBL-TE noise problem remain limited, with few examples beyond those presented in our preliminary work~\citep{demange_wavepackets_2024}. 

However, modal decomposition approaches have been extensively applied in the field of jet noise research, utilising a decomposition along the azimuthal direction, since the study of~\cite{cavalieri_axisymmetric_2012}. These methods have enabled the identification of a variety of flow mechanisms underlying jet noise production, supported by linear analyses ~\citep{colonius_parabolized_2010,karbanEmpiricalModelNoise2023,bugeatAcousticResolventAnalysis2024}.

The main motivation for the present work is to 
conduct experiments that explore the connection between SPFs and TE noise within the spanwise wavenumber domain. The aforementioned numerical studies on modal analysis for TBL-TE noise rely on simulations with spanwise periodicity, for which a spanwise Fourier decomposition is straightforward. This is not the case in experiments due to end conditions related to the test section. Our aim is therefore to reconcile earlier experimental results reporting short coherence lengths across the span with recent numerical work suggesting that low wavenumbers across the span (and hence long wavelengths) are responsible for TE noise. 
It should be noted that this work is accompanied by a companion paper ~\citep{Yuan2024arxiv}, which contains a detailed numerical simulation of one of the present experiments. In this work, a detailed modal analysis of the turbulent flow is performed, showing organised coherent structures that actually take the form of streamwise-elongated wavepackets in the turbulent boundary layer. Overall, these works are expected to lay the foundation for future modelling and control of TE noise as a phenomenon driven by spanwise organised motion in the turbulent flow. 

The structure of this paper is as follow: \S\,\ref{sec:toy-model} discusses the contradiction of spanwise length scales related to TE noise in modal and classical TE noise approaches. Subsequently, the experimental setup is presented in \S\,\ref{sec:exp_setup}, including the wind tunnel, airfoil models and instrumentation used. The post-processing of microphones signals is then discussed in \S\,\ref{sec:Processing_of_microphone_signals}. The results are divided into two sections. The first, \S\,\ref{sec:Results_TE_noise}, is focused on the characterisation of the acoustic field, in particular its spanwise length scales. The second, \S\,\ref{sec:correlation_SPF_acoustics}, examines the relation between the acoustic field and spanwise coherent SPFs and describes the SPFs spectra. Finally, \S\,\ref{sec:conclusions} presents the conclusions and perspectives of this work.

\section{Preliminary considerations on the relation between spanwise wavenumbers and coherence length}\label{sec:toy-model}

This section addresses the apparent contradiction between the two perspectives on the relevant spanwise length scales of flow structures driving broadband TE noise: i) the spanwise coherence length $\Lambda_z$, derived from second-order statistics, which considers the superposition of all fluctuations in the boundary layer; ii) the radiating spanwise wavelength $\lambda_z^{*}$ satisfying the scattering condition from~\cite{Nogueira2017}, which relies on a modal analysis of the SPFs to single out the radiating structures in the flow. The outcomes of this comparison are particularly important for the physical interpretation of flow structures driving TE noise, and to the choice of sensor placement in experiments and domain size in numerical simulations. 

We first illustrate the difference of scale between $\Lambda_z$ and $\lambda_z^{*}$ by considering the typical configurations studied in the TE noise literature: NACA0012 airfoils at low angles of attack ($\alpha \leq 6^\circ$) and Reynolds number of the order of $Re\approx10^6$. In these conditions, peak acoustic emissions associated with TE noise typically occur within $St\in[1,\, 20]$~\citep{brooks_trailing_1981,herrig_broadband_2013,herrBroadbandTrailingEdgeNoise2015}, when the Strouhal number $St=f c/u_\infty$ is based on the freestream velocity $u_\infty$ and chord length $c$. For this range, $\Lambda_z$ has been shown to be at most on the order of $\delta_{99}$ at the trailing edge~\citep{herrig_broadband_2013}.

In contrast, the scattering condition developed by~\cite{Nogueira2017} describes sound-producing SPFs as Fourier modes along the spanwise direction, with radiating wavelengths typically much larger than $\delta_{99}$. Using a modal ansatz for SPFs in Lighthill’s acoustic analogy with a tailored Green’s function, the scattering condition suggests that only modes with spanwise wavenumbers corresponding to supersonic phase velocities, $k_{\text{z}}^*$, can radiate sound, given by
\begin{equation}\label{eq:propagation}
k_{\text{z}}^{*} < k_s \text{,} 
\end{equation}
where $k_s=2\pi/\lambda_s$ is the acoustic wavenumber and $\lambda_s$ is the acoustic wavelength. This implies that the maximum radiating spanwise wavenumber, non-dimensional with $c$, is defined by the Helmholtz number based on the airfoil chord length
\begin{equation}\label{eq:He}
    \max(k_z^{*})c = k_s c = 2\pi St M = He\text{,}
\end{equation}
\noindent where $M=u_{\infty}/a_s$ is the Mach number with $a_s$ the speed of sound. 

For the maximum frequency in the TE noise range ($St=20$), which corresponds to $He=10$ for the Mach number considered in this study, this scattering condition yields a minimum radiating wavelength of $\min(\lambda_z^*)=0.63c$, or more than half the chord length. Thus, only spanwise structures with $\lambda_z^*\gg \delta_{99} \approx \Lambda_z$ contribute to TE noise at this frequency, indicating that the trailing edge acts as a low-pass filter of the SPFs’ spanwise wavenumbers. The main output of this analysis is that not all turbulent structures are expected to participate to TE noise.

This filtering effect also reconciles the apparent contradiction: the coherence length $\Lambda_z$ includes overlapping contributions from all turbulent scales, whether radiating or not. Although it may be used to determine statistically independent sources across the span~\citep{katoNumericalPredictionAerodynamic1993}, it may also obscure the specific radiating structures in the turbulent flow. Furthermore, as shown in \S\ref{sec:coherence_synthetic}, the coherence length can appear relatively small even when large-scale structures are present in the flow.

Consequently, a spanwise modal decomposition offers a way to isolate and examine the specific structures responsible for TE noise radiation. However, this approach requires sufficiently large spanwise domain sizes, $L_z$, to resolve these structures. The following sections aim to experimentally validate that large structures, resulting in low $k_z$ SPFs, correlate with distinct spanwise components of the acoustic field generated by an airfoil in flow.

\section{Experimental setup}\label{sec:exp_setup}

The aeroacoustic measurements were conducted in the acoustic wind tunnel of the TU Berlin, using two airfoil models with distinct instrumentation. This section begins with an introduction to the experimental facility and airfoil investigated. It then proceeds to discuss the specific setups employed to characterise the acoustic field and the surface pressure fluctuations near the trailing edge of the airfoil, before concluding with an overview of the experimental conditions investigated.  

\subsection{Aeroacoustic wind tunnel and model}

The experimental setup consisted of a rectangular nozzle releasing a free jet into a large anechoic chamber. The chamber has a volume of \SI{830}{\cubic\metre} and a lower cut-off frequency of \SI{63}{\hertz}. More details on the facility can be found in~\cite{schneehagen_design_2021}, although for another nozzle than the one used to produce the present results, with a similar shape and contraction ratio.

\begin{figure}
\centering
\begin{subfigure}{.5\textwidth}
    \centering
    \includegraphics[width=\columnwidth]{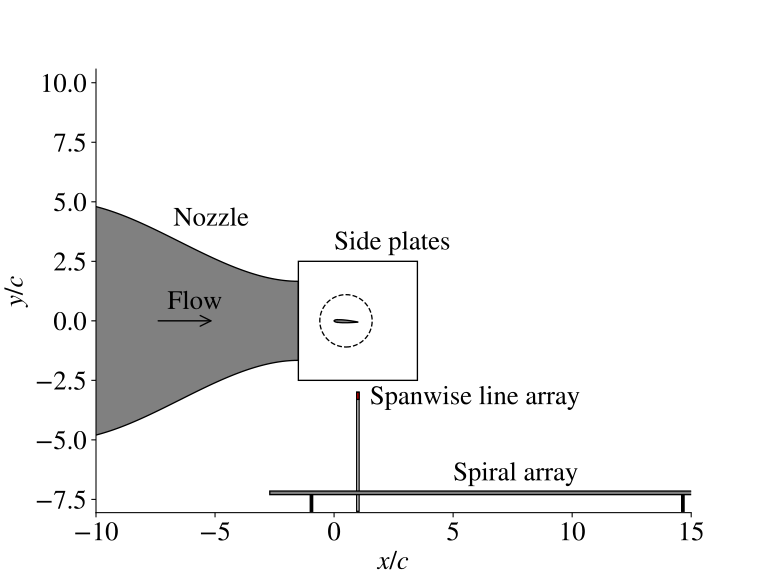}
    \caption{Side view of the experimental setup.}
    \label{fig:side_view_setup}
\end{subfigure}
\hfill
\begin{subfigure}{.45\textwidth}
    \centering
    \includegraphics[width=\columnwidth]{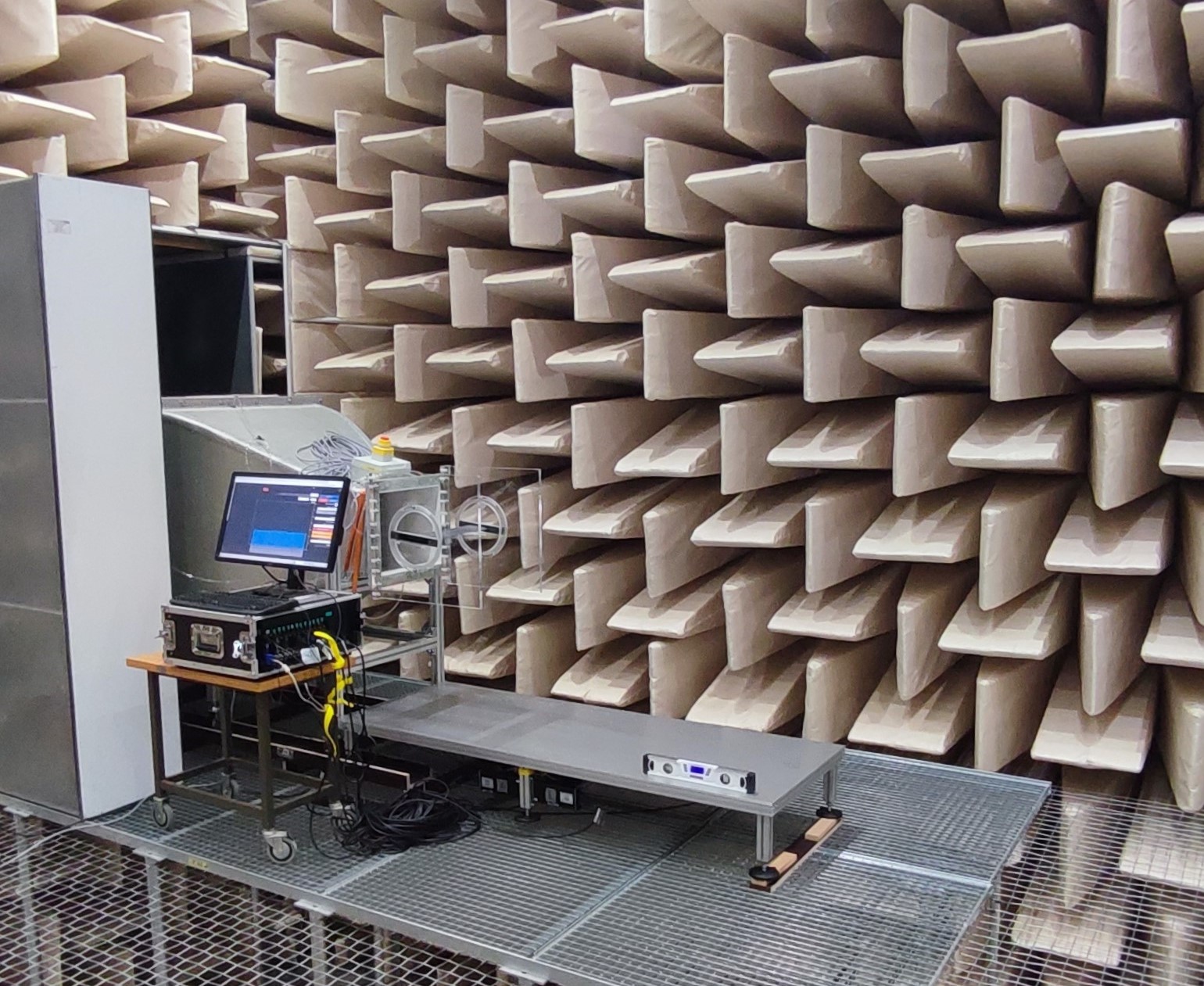}
    \caption{Picture of the experimental setup.}
    \label{fig:IMG_20220914_143230}
\end{subfigure}

\begin{subfigure}{.49\textwidth}
    \centering
    \includegraphics[width=\columnwidth]{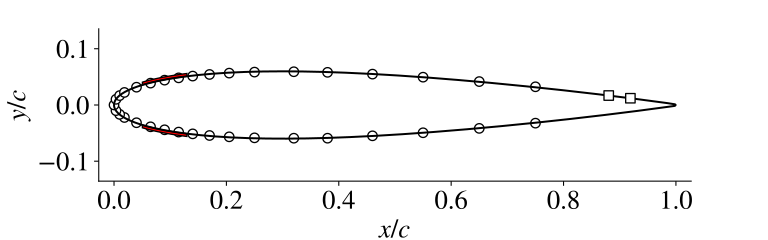}
    \caption{NACA0012 profile.}
    \label{fig:naca0012_sensors}
\end{subfigure}
\hfill
\begin{subfigure}{.49\textwidth}
    \centering
    \includegraphics[width=\columnwidth]{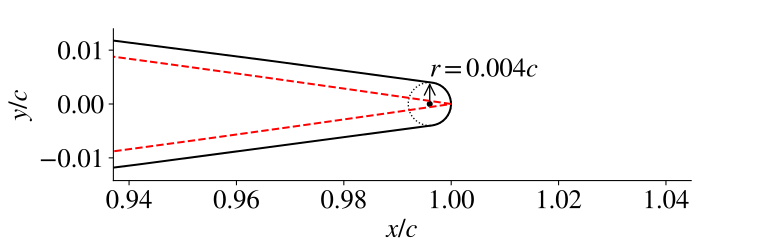} 
    \caption{Rounded trailing edge.}
    \label{fig:rounded_TE}
\end{subfigure}
\caption{Details of the experimental campaign with a sketch (a) and picture (b) of the setup, note that the spanwise line array is not installed in (b); and details of the profile (c) used in experiments with position of static ($\circ$) and dynamic ($\opensquare$) pressure sensors and trips (\textit{red}) and (d) geometry of the rounded trailing edge from experiments (\fullline) versus the classical profile ($\dashed$).}\label{fig:setup}
\end{figure}

A sketch and photograph of the setup are shown in figures~\ref{fig:side_view_setup} and~\ref{fig:IMG_20220914_143230}. Cartesian coordinates $(x,\,y,\,z)$ are used to denote the horizontal, vertical, and spanwise directions respectively. The nozzle has an exit section of height \SI{33}{\centi\metre} along $y$ and spanwise width \SI{40}{\centi\metre} along $z$. 

The airfoil was held between two circular inserts fitted into vertical side plates of length \SI{50}{\centi\metre} along $x$, mounted flush to the nozzle lips. The circular inserts allowed changing the geometric angle of attack of the profile, rotating around its centre. At a zero-degree geometric angle, the leading edge of the airfoil model was positioned \SI{15}{\centi\metre} downstream of the nozzle exit section, in its vertical midplane.

The airfoil models investigated in this work were two-dimensional NACA0012 profiles with a chord length of $c=$~\SI{10}{\centi\metre}, extruded along $z$ with a span equal to the width of the nozzle outlet, $L_{\text{z}}=$~\SI{40}{\centi\metre}. The profile was modified to include a rounded trailing edge of radius \SI{0.4}{\milli\metre}, similarly to other trailing-edge noise studies \citep{Ricciardi2022}. Both the airfoil and details of its trailing are shown in figures~\ref{fig:naca0012_sensors} and~\ref{fig:rounded_TE}.

Trailing-edge bluntness can cause significant shedding noise~\citep{Brooks1989}, which is not the phenomenon targeted in this study. We therefore performed a preliminary RANS of the 2D profile for conditions representative of this study and confirmed that our trailing edge radius was at least one order of magnitude smaller than the boundary-layer thickness at the trailing edge. This ensured that bluntness noise was not the dominant sound generating mechanism here.

The instrumented airfoil models were manufactured in-house using stereolithographic 3D printing with the \textit{formlabs} \textit{Rigid4K}\textsuperscript{\textcopyright} resin. This glass-filled resin allowed for a smooth finish and high stiffness of the model. Due to the limited volume of the 3D printer, the models were printed in five spanwise sections, held together by three traversing stainless steel spars and glued, resulting in the precise alignment of the sections and hardly detectable spanwise discontinuities of the profile.

roadband TE noise was achieved by tripping the boundary layer to turbulence. To this end, a zigzag tape (manufactured by Glasfaser Flugzeugservice) of measured height \SI{0.42}{\milli\metre}, angle $60^{\circ}$ and length \SI{8.3}{\milli\metre} was placed in streamwise direction on $x/c=0.05$ on both the pressure and suction sides. Based on the work of~\cite{DosSantos2022}, this rather large tripping with respect to the boundary-layer is expected to influence TE noise, although at frequencies higher than those of interest in this work. Furthermore, experimental parameters resulting in naturally broadband TE noise were investigated both with and without tripping to specifically investigate the influence of tripping in the modal analysis performed in this work. In addition, tonal trailing-edge noise was recorded when using a non-tripped airfoil for some operating conditions (see \S~\ref{sec:cases}), in consistency with previous experimental results using NACA0012 airfoils at similar Reynolds numbers~\citep{Lowson1994,Probsting2015_2}. 

Two airfoil models were produced for this work. The first was designed with inner channels for static pressure measurements, while the second featured an empty but closed volume near the trailing edge, which was used to house the PCB holding the micro-electro-mechanical systems (MEMS) for unsteady pressure measurements. Measurements of the static pressure distribution were mainly used to correct the angle of attack of the airfoil caused by the deflection of the open-jet flow when the model was set at a non-zero geometric angle~\citep{Brooks1984} $\alpha_{\text{geo}}\neq0^{\circ}$. The position of pressure ports is given in figure~\ref{fig:naca0012_sensors} and covers the first $75\%$ of the airfoil chord length. The effective angle of attack was determined by comparing the measured pressure coefficient ($c_p$) to the one returned by the panel-based method implemented in the XFOIL code~\citep{Drela1989}. To avoid uncertainties related to the position of the boundary-layer transition to turbulence, the  $c_p$  distribution was measured with the zigzag trip and the transition position was imposed in XFOIL. he following,  \enquote{$\alpha$} will refer to the effective (corrected) angle of attack. More details of the resulting correction are available in~\cite{demange_experimental_2023}. 

\subsection{Acoustic field measurements}\label{sec:acous_setup}

The acoustic field in the experiments was investigated using two different microphone arrays. A \textit{spiral array} in the horizontal plane positioned below the airfoil was used to identify sound sources employing the beamforming technique, and a \textit{line array} located under the trailing edge was used for the modal decomposition of the signals in the spanwise direction. The position of the microphone arrays in the setup is shown in figure~\ref{fig:side_view_setup}.

\begin{figure}
\centering
\begin{subfigure}{.6\textwidth}
    \centering
    \includegraphics[width=\columnwidth]{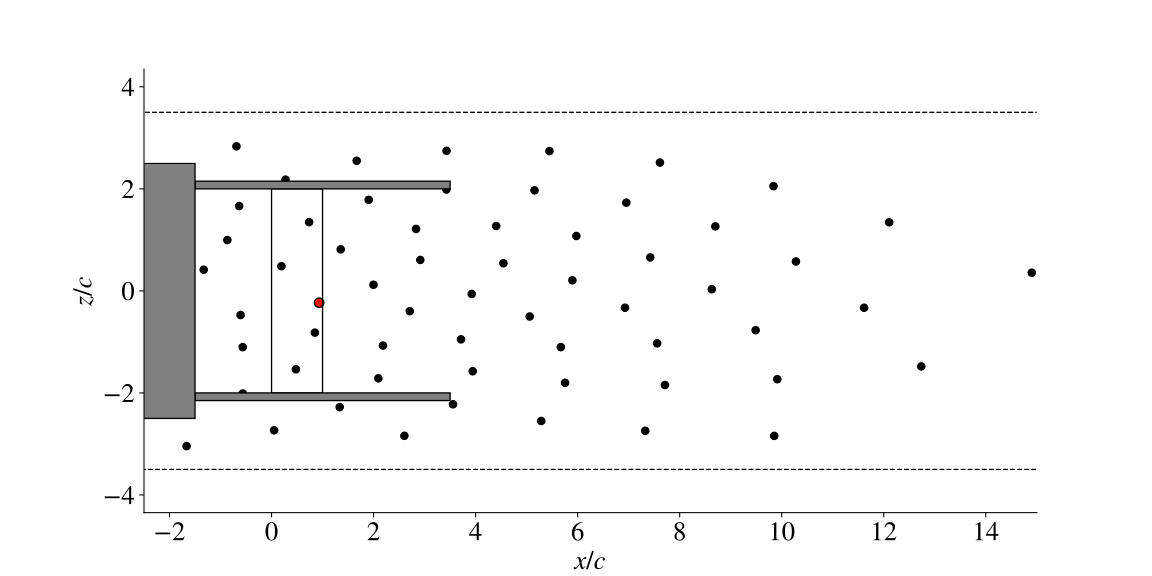}
    \caption{Spiral array.}
    \label{fig:elliptic_array}
\end{subfigure}
\begin{subfigure}{.3\textwidth}
    \centering
    \includegraphics[width=\columnwidth]{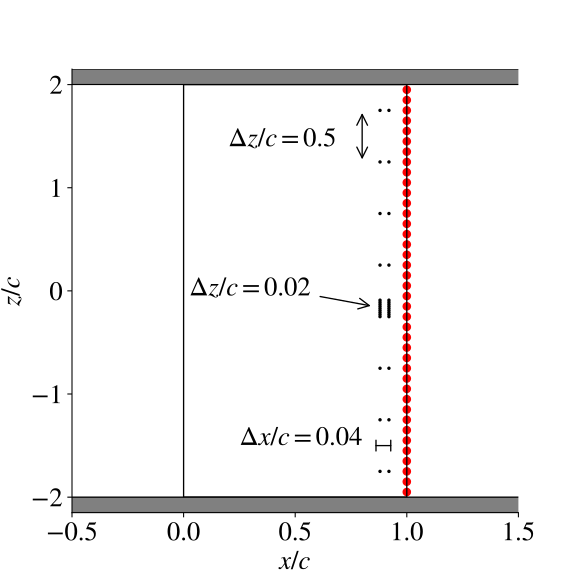} 
    \caption{MEMS and line array.}
    \label{fig:mems_line_arrays}
\end{subfigure}

\begin{subfigure}{.75\textwidth}
    \centering
    \includegraphics[width=\columnwidth]{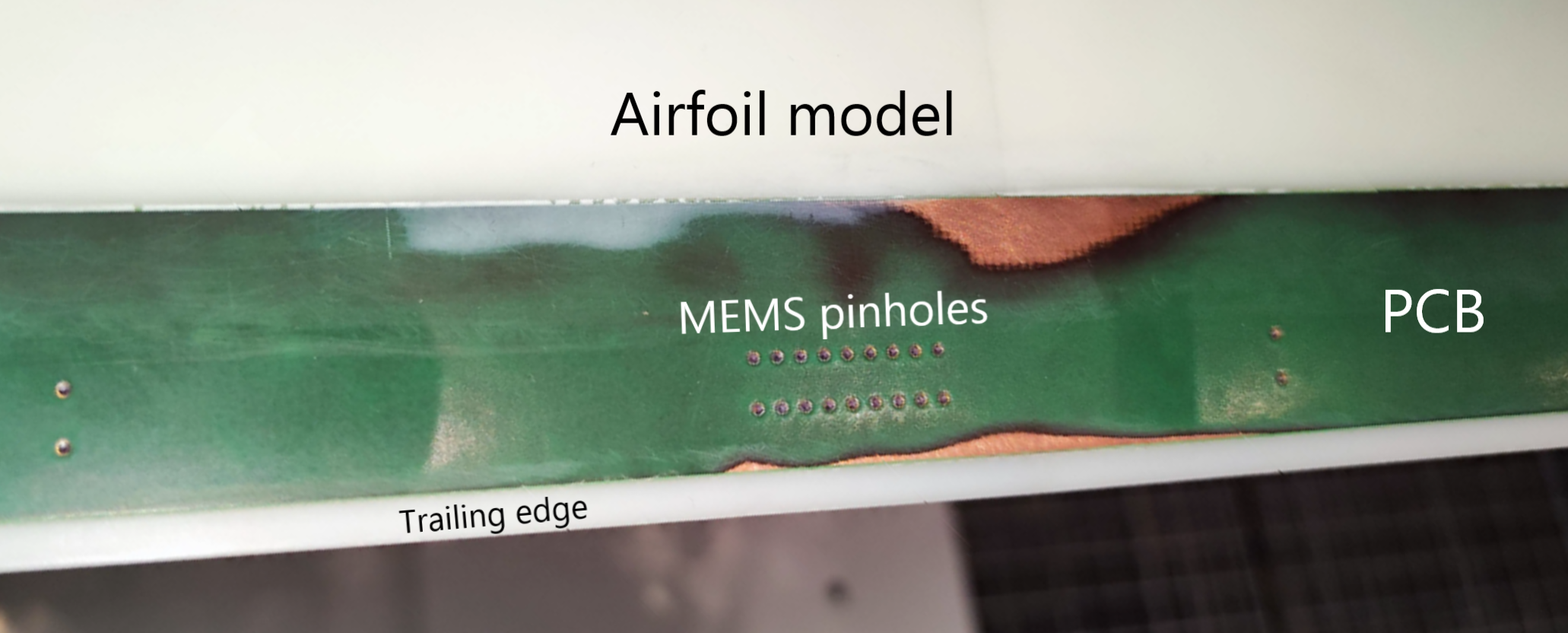} 
    \caption{Picture of the centre section of the airfoil model with the PCB holding the MEMS microphones.}
    \label{fig:airfoil_mems_installed_closeup}
\end{subfigure}
\caption{Top view of the microphone distribution for (a) the spiral array positioned $7.25c$ below the airfoil chord line and (b) the MEMS positioned on the surface of the airfoil (black dots) and the spanwise line array positioned $3c$ below the trailing edge (red dots); (c) close-up picture of the airfoil trailing edge with integrated PCB and MEMS microphones. Sanding the PCB flush with the profile exposed the first copper layer in the PCB without deteriorating the quality of the electric signals.}\label{fig:Arrays}
\end{figure}

The spiral array was positioned $7.25$ chord lengths below the centre of the airfoil at $\alpha_{\text{geo}}=0$~deg. It was composed of $64$ channels distributed in a sunflower pattern~\citep{Vogel1979} with parameters $H = 1.0$ and $V = 5.0$ according to~\citet{Sarradj2016}. The sensor distribution, shown in figure~\ref{fig:elliptic_array}, was scaled by $0.4$ in the spanwise direction to better fit the width of the nozzle exit section.

The line array was located three chord lengths below the trailing edge, for the airfoil at $\alpha_{\text{geo}}=0^{\circ}$. In this position, the microphones were outside the jet flow. The line array consisted of a horizontal spanwise line of $40$ microphones held in place by a 3D-printed support with a uniform spacing of \SI{10}{\milli\metre} from centre to centre. The position of the microphones in the horizontal plane is shown in figure~\ref{fig:mems_line_arrays}. To prevent electrical interference between the microphones, the support was held onto a rod that has been insulated with plastic tape. 

Both arrays consisted of \textsc{GRAS 40PL-1 Short CCP} microphones, in situ calibrated with a \textsc{Cirrus CR:515 class 1 calibrator} . Data were acquired with a Typhoon measurement system from Sinus Messtechnik GmbH, consisting of $12$ $8$-channel $24$-bit ADCs allowing for the synchronous measurement of signal from up to $96$ channels. The acquisition time was $t_{\text{meas}}=$~\SI{120}{\second} to ensure a converged coherence between sensors, with a frequency sampling of $f_s=$~\SI{51.2}{\kilo\hertz}.

\subsection{Surface pressure fluctuations measurements}\label{sec:MEMS_Setup}

To characterise the surface pressure fluctuations and to relate them to TE noise, the second airfoil model was equipped with an array of MEMS microphones near its trailing edge. The array consisted of two spanwise lines of $16$ sensors each, positioned at $x/c=88\%$ and $x/c=92\%$ respectively. Using two lines allows the prediction of the streamwise convection velocity of SPFs. The distribution of sensors along each line is illustrated in figure~\ref{fig:mems_line_arrays}. It features the following spanwise intervals (centre-to-centre): (i) $8$ sensors separated by $\Delta z =$~\SI{50}{\milli\metre}, targeting the measurement of pressure fluctuations with wavelengths spanning the whole airfoil width, and (ii) $8$ sensors separated by $\Delta z =$~\SI{2}{\milli\metre} (the minimum allowable with our setup) near the airfoil mid-span, targeting the measurement of pressure fluctuations that scale with the boundary layer thickness and the measurement of the spanwise coherence length.

The array was assembled by soldering TDK T4086 MEMS microphones to a flexible \SI{0.3}{\milli\metre}-thick printed circuit board (PCB). According to the manufacturer, these sensors yield a flat frequency response up to $f\approx$~\SI{10}{\kilo\hertz}, which was suitable for our setup. The PCB provided the power supply to the MEMS and was glued into a cavity prepared into the 3D-printed airfoil model. This way, the MEMS were positioned inside the airfoil with their sensor exposed to SPFs through \SI{0.2}{\milli\metre}-diameter pinholes drilled into the PCB. The PCB was then sanded to be flush with the surface of the airfoil model, as shown in figure~\ref{fig:airfoil_mems_installed_closeup}. 

TDK T4086 MEMS microphones include an output amplifier, such that the signals from the $32$ MEMS could be directly connected to the data acquisition system, allowing the synchronous measurement of SPFs with the line array of microphones for coherence calculations. Thus, the same acquisition time $t_{\text{meas}}=$~\SI{120}{\second} and sampling frequency $f_s=$~\SI{51.2}{\kilo\hertz} as for the acoustic array measurements were used for the MEMS measurements. During the campaign, the MEMS microphones were periodically calibrated in amplitude between runs using white noise from a loudspeaker against one of the acoustic microphones positioned near the trailing edge.

\subsection{Cases investigated}\label{sec:cases}

Seven experimental configurations were studied in this work, summarised in table~\ref{tab:main_parameters}. This enabled  consistency checks of the results across different flow parameters in the presence/absence of boundary layer tripping. The baseline case, $c2$, is highlighted in the table and is the one investigated in a high-fidelity simulation in part II~\citep{Yuan2024arxiv}. of this work.

\begin{table}
\centering
\begin{tabular}{l|ccccccc}
Cases & $\alpha~[\,^{\circ}\,]$ & $u_{\infty}$~[\,m.s$^{-1}$\,] & $T~[\,^{\circ}$C\,] & $Re/10^5$ & $M$ & trip & $\delta^*$/c\\
\hline
$c1$ & $0$ & $30$ & $16.0$ & $2$ & $0.088$ & \text{yes} & $0.64\%$ \\
\blue{$\mathbf{c2}$} & \blue{$\mathbf{3}$} & \blue{$\mathbf{30}$} & \blue{$\mathbf{16.0}$} & \blue{$\mathbf{2}$} & \blue{$\mathbf{0.088}$} & \blue{\textbf{yes}} & \blue{$\mathbf{0.86\%}$} \\
$c3$ & $6$ & $30$ & $16.0$ & $2$ & $0.088$ & \text{yes} & $1.25\%$ \\
$c4$ & $6$ & $30$ & $16.3$ & $2$ & $0.088$ & {\color{red} \text{no}} & $1.03\%$ \\ \hdashline
$c5$ & $6$ & $45,4$ & $16.4$ & $3$ & $0.133$ & {\color{red} \text{no}} & $0.91\%$ \\
$c6$ & $6$ & $45,4$ & $16.0$ & $3$ & $0.133$ & \text{yes} & $1.10\%$ \\
$c7$ & $3$ & $45,4$ & $16.0$ & $3$ & $0.133$ & \text{yes} & $0.97\%$ \\
\end{tabular}
\caption{Parameters of the cases investigated in the present experimental campaign. Non-dimensional numbers are given assuming dry air at $1$~\textit{atm}. The baseline case, \blue{$\mathbf{c2}$}, is the one investigated numerically in part. II of this work~\citep{Yuan2024arxiv}. The boundary layer displacement thickness at $x/c=92\%$, $\delta^*$, was obtained from XFOIL~\citep{Drela1989}.}\label{tab:main_parameters}
\end{table}

The experiments were conducted at two nozzle exit velocities, $u_{\infty}=$~\SI{30}{\metre\per\second} and $u_{\infty}=$~\SI{45.4}{\metre\per\second}, which correspond to chord-based Reynolds numbers of $Re=2\times10^5$ and $Re=3\times10^5$ and Mach numbers of $M=0.088$ and $M=0.133$ at room temperature $T=16^{\circ}$C, respectively. The undisturbed incoming flow velocity was measured using a Prandtl tube with a pressure transducer of range \SI{10}{\kPa} and accuracy of $0.1\%$ full span. Note that the Prandtl tube was removed from the flow during acoustic measurements to avoid its shedding noise. At $u_{\infty}=$~\SI{30}{\metre\per\second}, the average turbulence intensity at the  nozzle exit was measured to be  $Tu=0.2\%$~\citep{schneehagen_design_2021}.

Three values of the effective angle of attack, $\alpha=[0,\, 3,\, 6]^{\circ}$, were tested during measurements to investigate the influence of the airfoil incidence on the spanwise wavenumber content of TE noise and SPFs spectra. For $\alpha= 6^{\circ}$, the experiments yielded broadband TE noise without tripping for both nozzle exit velocities. This observation matches well with the tonal conditions reported by~\cite{Probsting2015_2} for a slightly lower level of inflow turbulence intensity of $Tu=0.1\%$. Therefore, we performed measurements both with and without the zigzag trip at this incidence in order to asses the influence of boundary-layer tripping (for both nozzle velocities).

\section{Processing of microphone signals}\label{sec:Processing_of_microphone_signals}

This section details the post-processing of signals from SPFs and acoustic microphones. Numerical implementation of these operations was done within the open-source code Acoular~\citep{sarradj_python_2017}.

\subsection{Frequency-domain operations}

The signals from microphones are analysed in frequency domain, by applying the Fourier transform of the temporal signal of the \emph{i-th} sensor, $\hat{s}_i(\omega)$, 
where $\omega=2\pi f$ and $f$ is the dimensional frequency. To access the relation of the phase and amplitude between the sensors, the cross spectral density (CSD) matrix is calculated using the Welch's method~\citep{Welch1967spectrum}. The signals from each sensor are first divided into $N_b$ segments of $N_{\text{FFT}}=4096$ samples with $50\%$ overlap, $s^{[k]}_i(t)$, before the Fourier transform is applied to obtain the complex coefficients, $\hat{s}^{[k]}_i(\omega)$. A Hanning window is used to limit spectral leakage. Entries of the CSD for the signals of the $i$-th and $j$-th sensors are then obtained by averaging the complex coefficients over all blocks
\begin{equation}\label{eq:csd}
	C_{ij}(\omega)=\frac{1}{N_{\text{b}}}\sum_{k=1}^{N_{\text{b}}}\hat{s}^{[k]}_i(\omega)\,\hat{s}^{*[k]}_j(\omega),
\end{equation}
where $\cdot^*$ denotes the conjugate transpose and the diagonal of $C_{ij}$ contains the auto-spectra, $P_{xx}$, from each microphone signal. The resulting frequency resolution is $\Delta f=$~\SI{12.5}{\hertz} or $\Delta {He}=0.023$ with this setup.

The power spectra are converted to sound pressure levels (SPLs) with respect to a reference sound pressure of $p_{0}=$~\SI{20}{\micro\pascal}, given in decibel (dB) as
\begin{equation}\label{eq:lp}
	SPL=10 \log_{10} \left(\frac{{\hat{s}}^{2}}{{p}_{0}^{2}}\right).
\end{equation}



When considering the signals from streamwise-adjacent pairs of MEMS microphones separated by $\Delta x$, the streamwise convection velocity of SPFs is estimated from the phase angle $\varphi_{xy}$ of the CSD between signals~\citep{brooks_trailing_1981} as 
\begin{equation}\label{eq:cph_x}
	c_{ph}(\omega) = \frac{\omega\Delta x}{\varphi_{xy}}.
\end{equation}

\subsection{Spanwise wavenumber decomposition}\label{sec_theory_span_decomposition}

Due to the presence of side plates, the present experimental setup deviates from an idealised infinite wing with a spanwise-homogeneous mean flow, for which a Fourier decomposition of the acoustic and SPF signal along $z$ would be straightforward. To circumvent this issue, we first investigate the cross spectral density matrix of the spanwise line array signals to investigate the spanwise organisation of the acoustic field. More specifically, we consider CSD entries corresponding to the correlation between the centre line sensor (mic. $20$) with all other sensors, $C_{20,j}$ for $j\in[0,\,40]$.  We then employ a spectral proper orthogonal decomposition (SPOD) of the corresponding entries of the CSD at the frequencies of interest, which isolates the most energetic time-span-coherent structures in the acoustic signals~\cite{lumey_stochastic_1970,berkooz_proper_1993,Schmidt2020} . The benefit of this approach is that it may reveal the organisation of the line signals, without assuming that the signal is periodic in the spanwise direction.
 
The SPOD method is based on an eigenvalue decomposition of the CSD-matrix~\eqref{eq:csd} entries corresponding to the product of signal from the centre sensor $(i=20)$ with that of all other sensors $(j\in[0,\,40])$,
\begin{equation}\label{eq:eig_csd}
	C_{i=20,\,j\in[0,\,40]}(\omega)\hat{\Psi}(\omega) = \sigma(\omega)\hat{\Psi}(\omega),
\end{equation}
where the diagonal of $\sigma$ represents the real-valued eigenvalues and $\hat{\Psi}$ the orthogonal eigenvectors of the CSD. The eigenvalues represent the energy contained in the structures as a function of frequency, and the eigenvectors the corresponding shapes $\hat{\Psi}_i$ along the line.

The SPOD analysis, described in \S~\ref{sec:acous_charac}, showed that the SPOD modes are very similar to spanwise Fourier modes. This suggests that a Fourier basis is a suitable representation of the data, even for the finite-span line array data considered here. Hence, we can safely apply a spanwise Fourier transform of the signals from the acoustic line array, defined as
\begin{equation}\label{eq:dft_z}
    \hat{s}(k_z, t) = \int_{-L_{z}/2}^{L_{z}/2} s(z, t)e^{-ik_z z}dz,
\end{equation}
where $k_{\text{z}}$ is the spanwise wavenumber.

The spanwise array used in this work is composed of $N_z=40$ microphones separated by $\Delta z=$~\SI{0.01}{\metre} and has a length of $L_{\text{z}}=$~\SI{0.4}{\metre} This setup resolves non-dimensional wavenumbers within $k_{\text{z}}c\in[-31.4,\,29.8]$~\si{\metre}$^{-1}$ with a resolution of $\Delta k_{\text{z}}c=1.57$. Note that this range of $k_{\text{z}}$ is compatible with the sound-scattering wavenumbers obtained from the TE scattering condition up to $He=10$, as detailed in \S~\ref{sec:toy-model}. These wavenumbers correspond to positive non-dimensional wavelengths within $\lambda_z/c\in[0.2,\,4]$. 


\subsection{Acoustic beamforming}

Although experiments are conducted in an anechoic wind tunnel, it is anticipated that a number of unintended phenomena besides TE noise will contribute to the overall sound field captured by the microphones. These can include leading edge noise induced by inflow turbulence, corner noise at the junction of the side plates and the model leading edge, or vortex shedding at the nozzle upper/lower lips and at the downstream edge of the side plates. In this study, we utilise the conventional delay-and-sum beamforming approach in the frequency domain to localise the acoustic sources in the setup using the signals of the spiral array. 

For the present setup, the focus plane of the beamforming is the two-dimensional horizontal plane passing by the trailing edge at $\alpha=0^{\circ}$, at $y=0$, which is divided into square cells of \SI{1}{\centi\metre} width. The focus area is defined within $-2.75 \leq x/c \leq 7.35$ in the streamwise direction and $-4 \leq z/c \leq 4$ in the spanwise direction.

In practice, the beamforming formulation is based on the phase and amplitude differences information contained in the CSD matrix~\eqref{eq:csd}, where the diagonal entry has been removed beforehand since it may contain uncorrelated self-noise of the channels~\citep{herold_performance_2017}. The beamforming output is the squared sound pressure contribution at the array centre from each of the considered source points $x_t$ in the discretized focus plane, obtained from
\begin{equation}\label{eq:bf}
    b(x_{t}) = h_{i}^{*}\left(x_{t}\right) \, C_{ij} \, h_{j}\left(x_{t}\right),
\end{equation} 
where $h$ is the steering vector containing the phase shift and amplitude correction according to the transfer functions from the points in the focus plane to the microphone positions.

For a given focus point $x_t$, the steering vector is defined assuming a monopole sound propagation model~\citep{sarradj_three-dimensional_2012} by
\begin{equation}\label{eq:steering_vec}
    h_{i} = \frac{1}{r_{s,0} \, r_{s,i} \, \sum_{l=1}^{N}{r_{s,l}}^{-2}}\,e^{-j k_{0}(r_{s,i}-r_{s,0})},
\end{equation} 
where $N=64$ is the number of microphones in the spiral array, $k_s$ is the acoustic wavenumber, $r_{s,i}$ is the distance between the $i$-th microphone and the focus point, and $r_{s,0}$ is the distance from the focus point and a reference point at the centre of the array. The convection of acoustic waves within the free jet is accounted for by using a ray-casting approach~\citep{sarradjFastRayCasting2017}, based on an analytical approximation of the flow field out of a slot jet.

Depending on the array geometry and distribution of sound sources, conventional beamforming is known to produce spatial artefacts, e.g. side lobes, that can be removed with deconvolution techniques such as \textsc{DAMAS}~\citep{brooks_deconvolution_2006} or \textsc{CLEAN-SC}~\citep{sijtsma_clean_2007}. However, only the conventional beamforming method is used in this work since the focus of our investigation are coherent sound sources: SPFs with Fourier modes shapes along the span.
\section{Identification and spanwise decomposition of TE noise}\label{sec:Results_TE_noise}

\subsection{Localisation of sound sources using beamforming}\label{sec:results_beamforming}

The objective of this section is to identify the frequency  range for which TE noise is the primary contributor of the radiated acoustics. This range will then be the focus of the  subsequent investigations.

\begin{figure}
    \begin{subfigure}{.49\textwidth}
        \centering
        \includegraphics[width=\columnwidth]{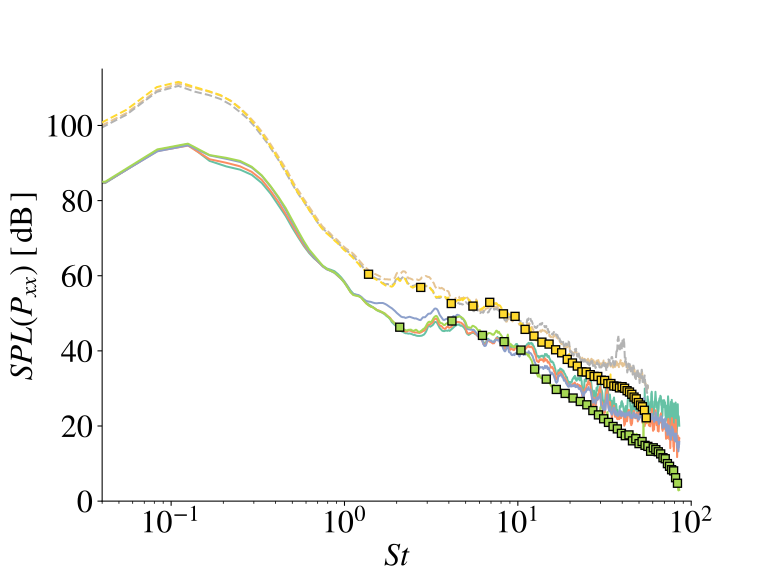}
        \caption{Without scaling.}
        \label{fig:SPL_noScaling_singleLinemic_allcases}
    \end{subfigure}
    \begin{subfigure}{.49\textwidth}
        \centering
        \includegraphics[width=\columnwidth]{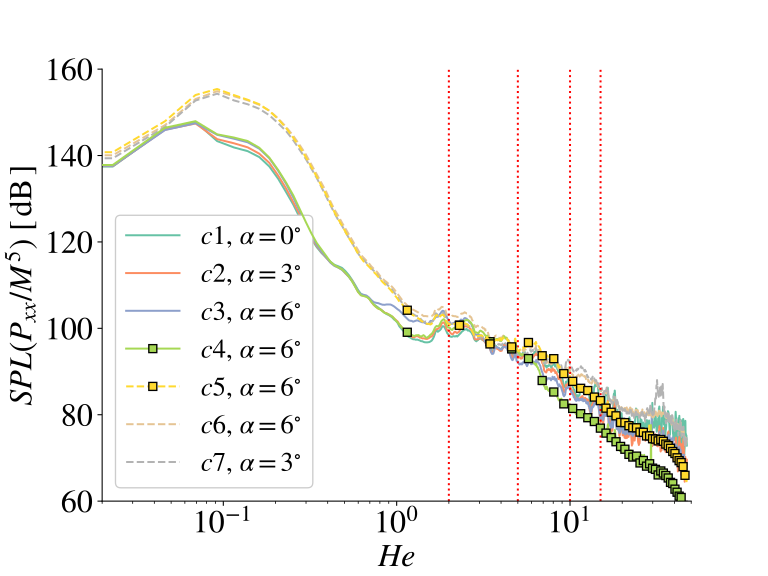} 
        \caption{With $M^5$ scaling.}
        \label{fig:SPL_M5Scaling_singleLinemic_allcases}
    \end{subfigure}
    
    \caption{Sound pressure levels scaled with $M^5$ from a single microphone at the centre of the line array for all seven experimental conditions. The effect of freestream velocity is highlighted by differentiating the cases with $M=0.08$ and $Re=2\times10^5$ (\fullline) from those with $M=0.133$ and $Re=3\times10^5$ (\dashed). Non-tripped cases are highlighted by symbols (\fullsquare). Vertical dotted lines in (b) indicate the frequencies for which beamforming results are shown in figure~\ref{fig:beamforming}.}\label{fig:SPL_line_single_allcases}
\end{figure}

We first consider the sound-pressure level spectra from a single-microphone for  all experimental conditions. It is shown in figures~\ref{fig:SPL_noScaling_singleLinemic_allcases} and~\ref{fig:SPL_M5Scaling_singleLinemic_allcases} as a function of the Strouhal and Helmholtz numbers, respectively. All spectra display similar features, including a large amplitude hump at very low frequency, a second broad hump at mid-frequency range, and small-amplitude sharp peaks at high frequencies.

The spectra shown in figure~\ref{fig:SPL_M5Scaling_singleLinemic_allcases} are scaled with $M^5$. This is motivated by the early work of~\cite{powell_aerodynamic_1959} who identified TE noise to scale with the freestream velocity raised to the fifth power.  Using this scaling we observe good collapse for $2 \leq He \leq 7$, with the peak of the mid-frequency hump located at $He\approx2$, which indicates that this hump is likely associated with TE noise. Therefore, we focus the beamforming investigation on this frequency range. We further observe that smaller-amplitude peaks from the spectra of all cases collapse along the frequency axis for $He\geq1$, while an offset is observed between cases at different freestream velocities when using $St$.

On the contrary, the SPLs of the low-frequency peak do not collapse when using the $M^5$ scaling when plotted against the Helmholtz number. However, a collapse is found when plotted against Strouhal number, with the peak at around $St=0.1$.  Based on the different scaling behaviour, we  expect that the low-frequency peak is not related to TE noise. A detailed investigation of this range is, whoever, difficult using the beam forming technique. 

\begin{figure}
    \begin{subfigure}{.49\textwidth}
        \centering
        \includegraphics[width=\columnwidth]{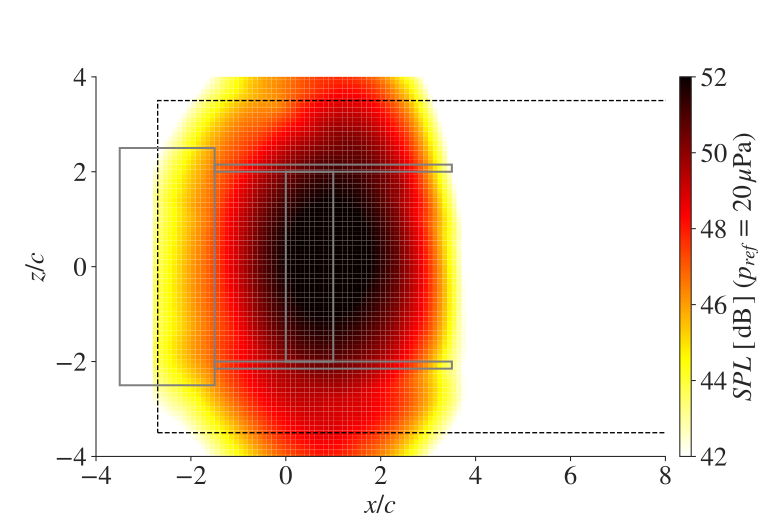}
        \caption{$He=2$, $St = 3.6$}
        \label{fig:BF_u30aoa4d8trip_He2}
    \end{subfigure}
    \begin{subfigure}{.49\textwidth}
        \centering
        \includegraphics[width=\columnwidth]{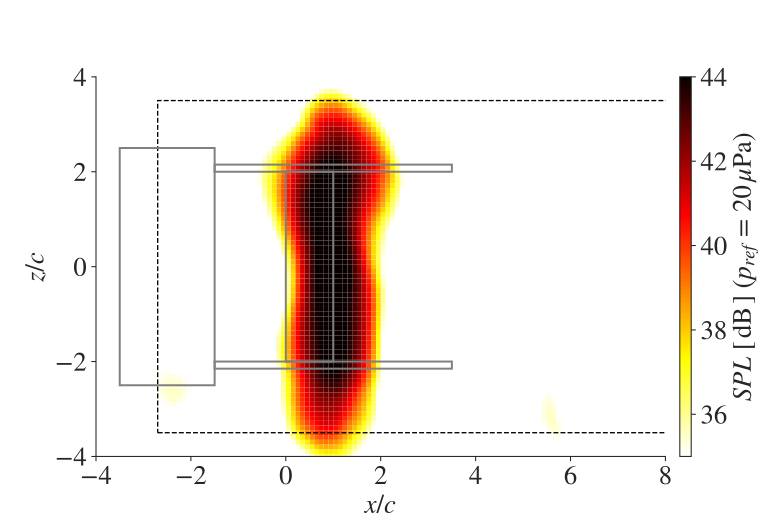} 
        \caption{$He=5$, $St = 9$}
        \label{fig:BF_u30aoa4d8trip_He5}
    \end{subfigure}

    \begin{subfigure}{.49\textwidth}
        \centering
        \includegraphics[width=\columnwidth]{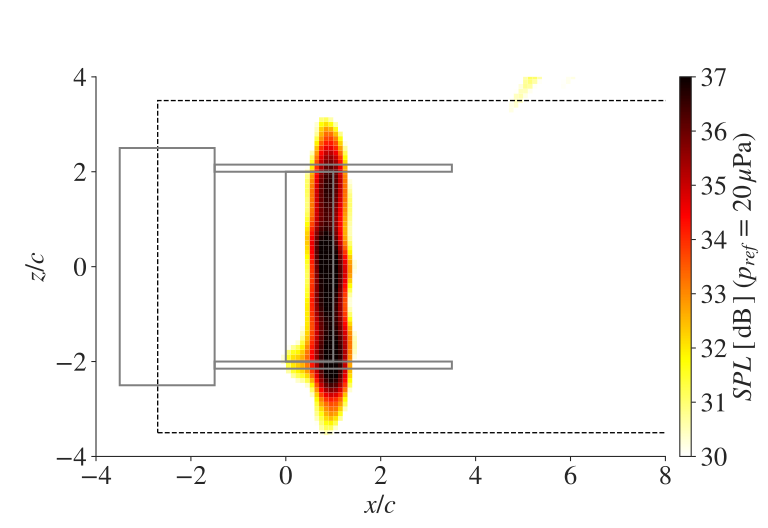}
        \caption{$He=10$, $St = 18.1$}
        \label{fig:BF_u30aoa4d8trip_He7}
    \end{subfigure}
    \begin{subfigure}{.49\textwidth}
        \centering
        \includegraphics[width=\columnwidth]{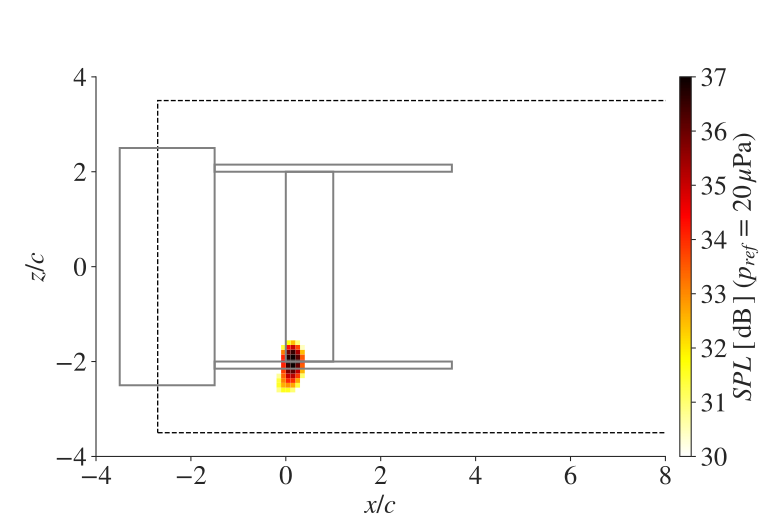} 
        \caption{$He=15$, $St = 27.1$}
        \label{fig:BF_u30aoa4d8trip_He15}
    \end{subfigure}
    
    \caption{One octave source maps for the baseline case ($Re=2\times10^5$, $\alpha=3^{\circ}$, \textit{tripped}) around (a) $He=2$, (b) $He=5$, (c) $He=10$, and (d) $He=15$, using the classical beamforming method. Contours of SPLs within $20\%$ of the maximum value are shown for each frequency.}\label{fig:beamforming}
\end{figure}

Beamforming is applied to the signals recorded with the spiral array microphones for $He=[2,\,5,\,10,\,15]$, as indicated by dotted vertical lines in figure~\ref{fig:SPL_M5Scaling_singleLinemic_allcases}. The resulting sound maps are shown in figure~\ref{fig:beamforming} for the baseline ($c2$) case. For the three lower Helmholtz numbers, the dominant sound source is found at the trailing edge of the airfoil, confirming that TE noise is the dominant contributor of the emitted sound field at this frequency range. As expected, results are less precise for the lowest frequency, since phase differences between microphones decrease with $He$~\citep{herold_performance_2017}. Nonetheless, the contours are still clearly centred around the trailing edge in this case.

In figure~\ref{fig:BF_u30aoa4d8trip_He15}, we observed that for $He=15$ the dominant sound source is located at the corner between the airfoil leading edge and one of the side plates. Similar undesired corner or junction leading edge sound sources have been reported in previous experiments investigating airfoil leading~\citep{zamponi_role_2020,hales_mathematical_2023} and trailing~\citep{oerlemans_aeroacoustic_2004} edge noise. This sound source is irrelevant for the integrated spectra from specific regions of the focus plane using beamforming, but it might pollute the results from the line array measurements performed in this work. Therefore, the frequency range where corner sound is dominant is avoided altogether. For all cases, the corner source only becomes predominant at $He>10$.

Summarizing the beamforming and single microphone results, the common frequency range where TE is the main contributor to the sound field in all considered cases is $1 \leq He \leq 10$. This range translates to $1.8 \leq St \leq 18$ at $M=0.088$ and $1.2 \leq St \leq 12$ at $M=0.133$. It agrees very well with the range reported in ~\cite{sanders_trailing-edge_2022} who compared the  NACA0012 TE noise spectra from several literature sources.

\subsection{Characterisation of the acoustic field}\label{sec:acous_charac}

Now that the frequency range of TE noise is determined, this section characterises the spanwise organisation of the acoustic field based on measurements from the line array. As mentioned in section~\ref{sec_theory_span_decomposition}, the first analysis of the line signals does not decompose the array signals in Fourier modes, but relies on SPOD to extract the most energetic, in the sense of squared pressure fluctuations, span-time coherent structures of the acoustic field.

\begin{figure}
    \centering
    \begin{subfigure}{\textwidth}
        \centering
        \includegraphics[width=\textwidth]{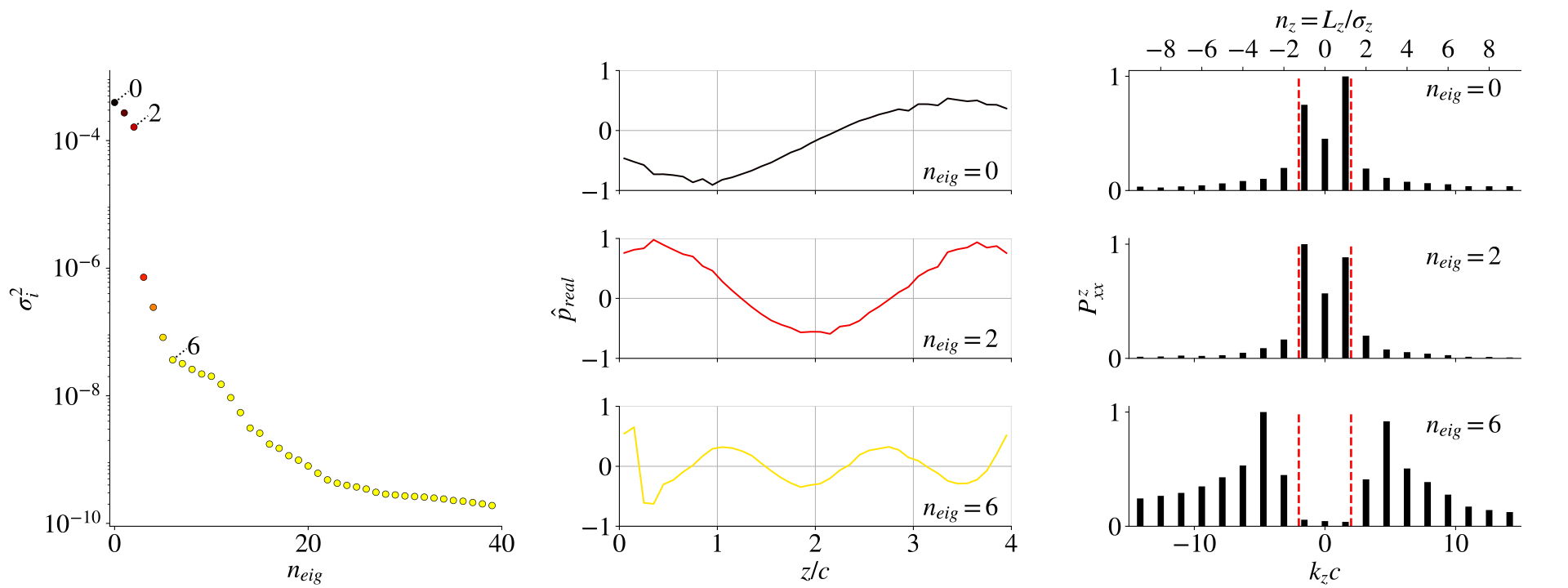}
        \caption{$He =2$}
        \label{fig:SPODLine_c2_He2}
    \end{subfigure}
    
    \begin{subfigure}{\textwidth}
        \centering
        \includegraphics[width=\textwidth]{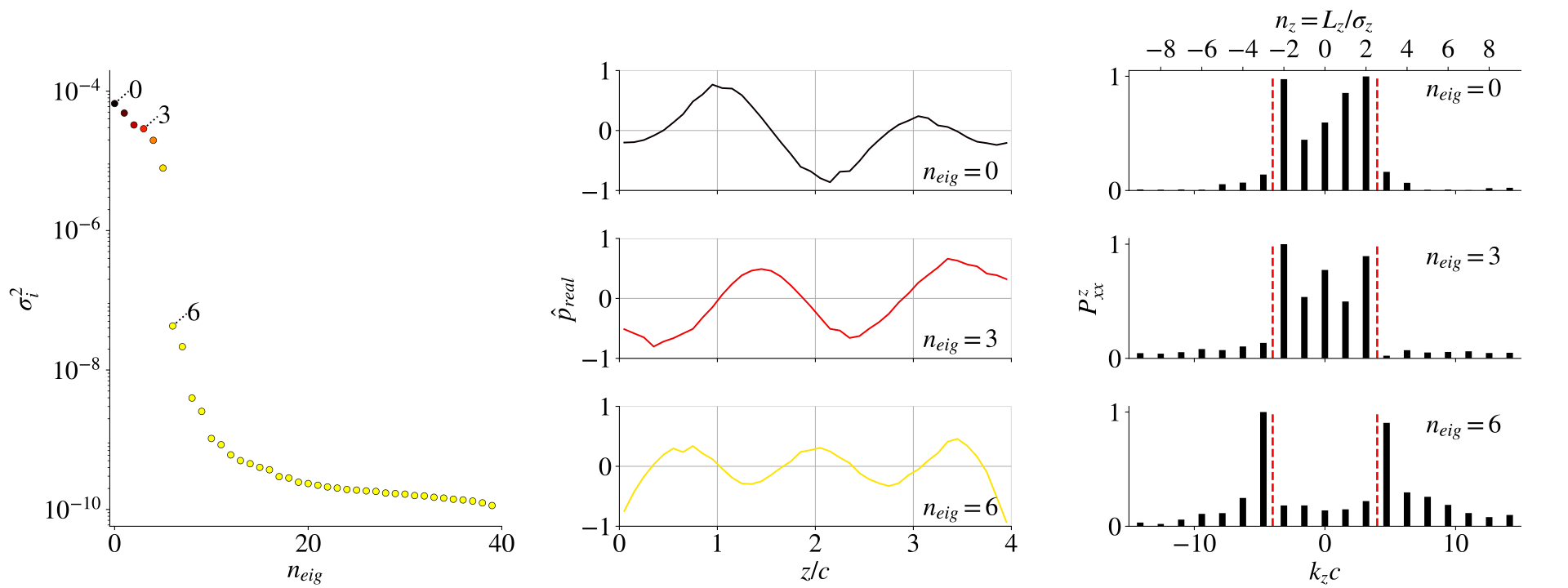}
        \caption{$He=4$}
        \label{fig:SPODLine_c2_He4}
    \end{subfigure}

    \begin{subfigure}{\textwidth}
        \centering
        \includegraphics[width=\textwidth]{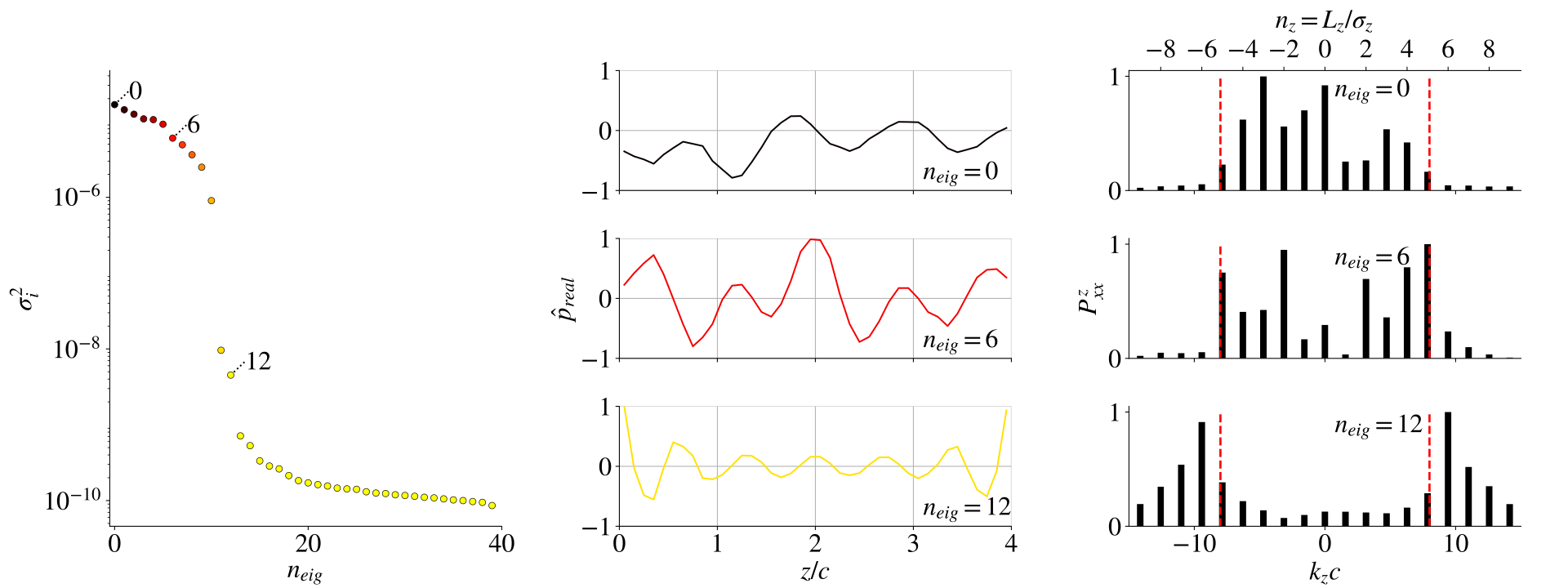} 
        \caption{ $He=8$}
        \label{fig:SPODLine_c2_He8}
    \end{subfigure}
    \caption{SPOD decomposition of the line-array signals for the baseline case ($Re = 2\times10^5$, $\alpha=3^{\circ}$, \textit{trip}). Results show the SPOD spectrum (\textit{left}), the real part of selected modes (\textit{middle}), and the PSD from the spanwise Fourier transform of the complex mode shapes (\textit{right}) together with the acoustic wavenumber, $k_s$ ({\color{Red} $\dashed$}).}\label{fig:SPODLine_c2}
\end{figure}

Results of the SPOD analysis are shown in figure~\ref{fig:SPODLine_c2} for the baseline case, $\mathbf{c2}$, at three Helmholtz numbers $He=[2,\,4,\,8]$. Results from the baseline case are representative of those obtained with the other experimental conditions (not shown). For all frequencies, the eigenvalue spectra (presented in the \textit{left column}) indicate a low-rank behaviour of the pressure fluctuations. In the case of $He=2$, the three leading eigenvalues account for over $99\%$ of the total energy, while the energy associated with the subsequent sub-leading modes shows a sudden decline of at least two orders of magnitude.
A comparable trend is observed at higher Helmholtz numbers, albeit with an increasing number of eigenvalues containing a significant proportion of the total energy. This indicates the presence of a frequency-dependent distribution of energy, which we now demonstrate is associated with the scattering condition.

The middle column of Figure~\ref{fig:SPODLine_c2} shows the real part of the SPOD pressure modes determined from the line signals. It offers insight into the low-rank behaviour observed in the SPOD spectrum. Three modes are shown at each frequency, indicated by their number in the energy spectra. Two observations can be made: (i) the SPOD mode shapes are very similar to Fourier modes, and (ii), modes corresponding to the dominant eigenvalues display lower wavelengths than that of eigenvalues below the energy drop. 

These observations can be related to the theoretical TE scattering condition~\citep{Nogueira2017} discussed in \S~\ref{sec:toy-model}, by examining the PSD of the spanwise Fourier transformed complex SPOD mode shapes, $P_{xx}^{z}$. The objective here is to relate the energy drop to the spanwise wavenumber of the SPOD modes. The resulting two-sided spectra (normalised for convenience) are shown in the \textit{right column} of figure~\ref{fig:SPODLine_c2}, and the acoustic wavenumber, $k_s$, is indicated with vertical dashed lines for each Helmholtz number $He$.

The $P_{xx}^{z}$-spectra show that the SPOD modes are not associated with a single spanwise wavenumber, $k_z$, but are likely the result of a superposition of left and right travelling acoustic waves, indicated by the presence of both positive and negative $k_z$ in the spectrum of each mode. Nonetheless, an excellent agreement is found between TE scattering condition and the wavenumber contents of SPOD modes: at each frequency, the dominant eigenvalues in terms of acoustic energy correspond to eigenvectors whose shape contains only $|k_z|<k_s$ fluctuations. 

\begin{figure}
    \centering
    \begin{subfigure}{.49\textwidth}
        \centering
        \includegraphics[width=\textwidth]{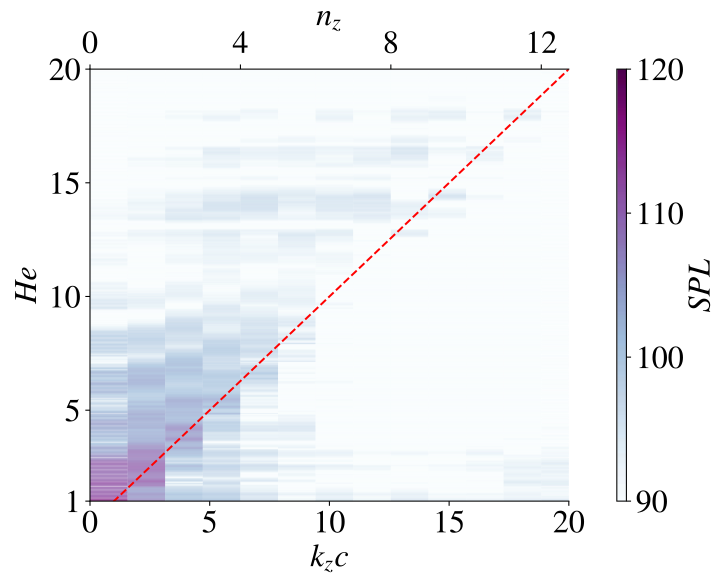}
        \caption{c1: $\alpha=0^\circ$, \textit{tripped}}
        \label{fig:scattering_condition_M5scale_c1}
    \end{subfigure}
    \hfill
    \begin{subfigure}{.49\textwidth}
        \centering
        \includegraphics[width=\textwidth]{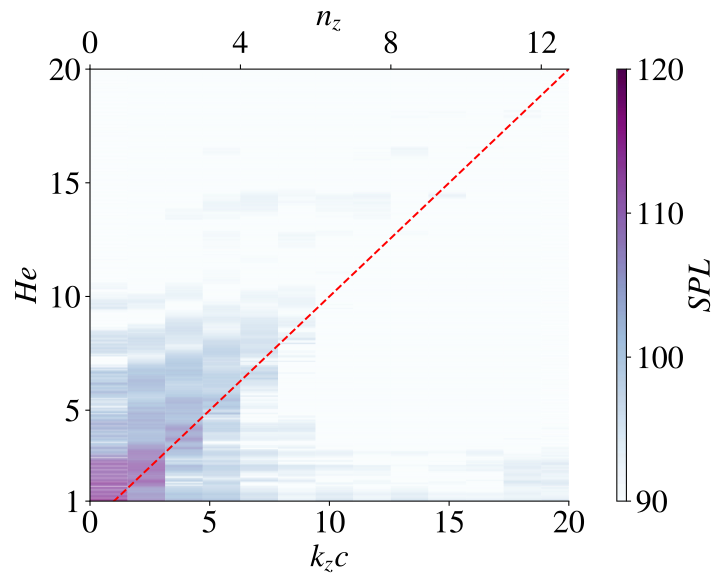} 
        \caption{c2: $\alpha=3^\circ$, \textit{tripped}}
        \label{fig:scattering_condition_M5scale_c2}
    \end{subfigure}
    
    \begin{subfigure}{.49\textwidth}
        \centering
        \includegraphics[width=\textwidth]{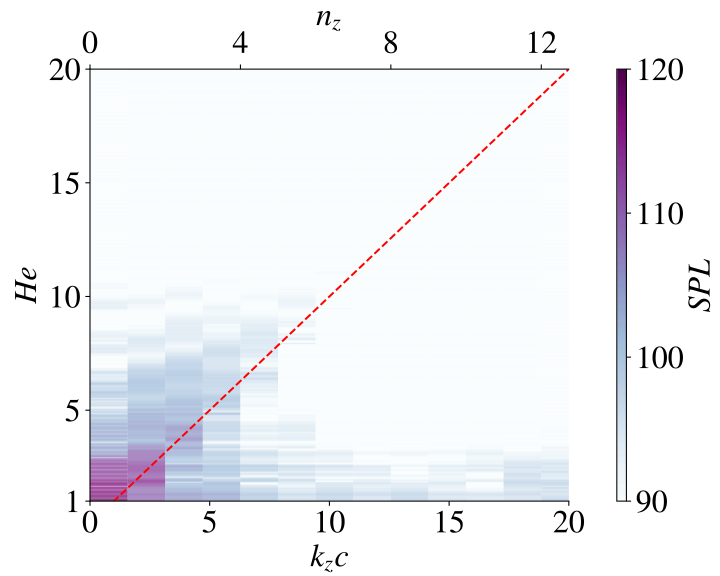} 
        \caption{c3: $\alpha=6^\circ$, \textit{tripped}}
        \label{fig:scattering_condition_M5scale_c3}
    \end{subfigure}
    \hfill
    \begin{subfigure}{.49\textwidth}
        \centering
        \includegraphics[width=\textwidth]{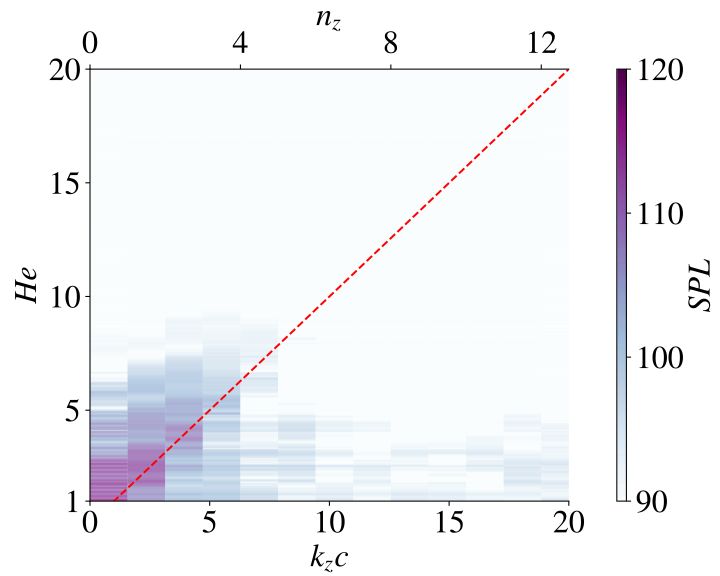} 
        \caption{c4: $\alpha=6^\circ$, \textit{clean}}
        \label{fig:scattering_condition_M5scale_c4}
    \end{subfigure}
    \caption{Sound pressure levels scaled by $M^5$ of the spanwise Fourier transform of the line array CSD, considering the centre microphone as reference, for tripped cases with $u_{\infty}=30$~m.s$^{-1}$ and increasing angles of attack (a) $\alpha=0^\circ$ deg, (b) $\alpha=3^\circ$ deg, and (c) $\alpha=6^\circ$, and (d) the clean airfoil at $\alpha=6^\circ$.}\label{fig:scattering_condition}
\end{figure}

As the SPOD shows that Fourier modes are a suitable representation of the acoustic field, we can directly apply the spanwise Fourier transform~\eqref{eq:dft_z} to the CSD matrix entries $C_{20,j}$. The resulting SPL spectra in the two-dimensional plane, $\left(k_z c,\,He\right)$, are shown in figure~\ref{fig:scattering_condition} for cases with $Re=2\times10^5$. Note that the spectra are plotted as continuous contours along $k_z$ for readability, but one should bear in mind that spectra are discrete along this axis. Moreover, only the positive wavenumbers are shown in the figures, since the spectra were approximately symmetric with respect to the $He$-axis.
For all cases, the spectra show that highest SPL is generally obtained for $\left(k_z c,\,He\right) = \left(0,\,1\right)$. This clearly shows that TE noise is predominantly generated by low-frequency fluctuations and very large spanwise wavelengths. 

In all cases, a good overall agreement is obtained between the data and the scattering condition: the highest SPLs are obtained for $k_z \leq k_s$, which corresponds to the upper left triangle of the plots. By comparing the spectra of the different cases, we observe that increasing angles of attack results in a faster decay of the SPL with increasing $He$ and $k_z c$. Further, it indicates that the tripping device results into slightly higher SPL at low frequency.

Some discrepancy with the theory is observed for all cases at low frequencies, $He \leq 5$, where significant SPLs are nevertheless observed for $k_z > k_s$. A first observation is that the scattering condition is related to far-field radiation; wavenumbers with $k_z\geq k_s$ lead to evanescent fluctuations which, despite their decay, may still be detected at the microphone position. Furthermore, we observe that applying a spatial window to the CSD entries prior to the spanwise Fourier transform resulted in a reduction of the SPL in this region of the spectrum, but also deteriorated the agreement with the scattering condition at low frequencies ($He\leq5$). This suggests the presence of spectral leakage in the analysis, which could be mitigated by increasing the airfoil span. Thus, the disagreement with the theory at $He \leq 5$ is likely related to the signal processing and spanwise width of the domain. Nevertheless, this observation constitutestisfying experimental validation of the scattering condition derived by~\cite{Nogueira2017} for airfoil TE noise. 

\begin{figure}
\centering
\includegraphics[width=.85\textwidth]{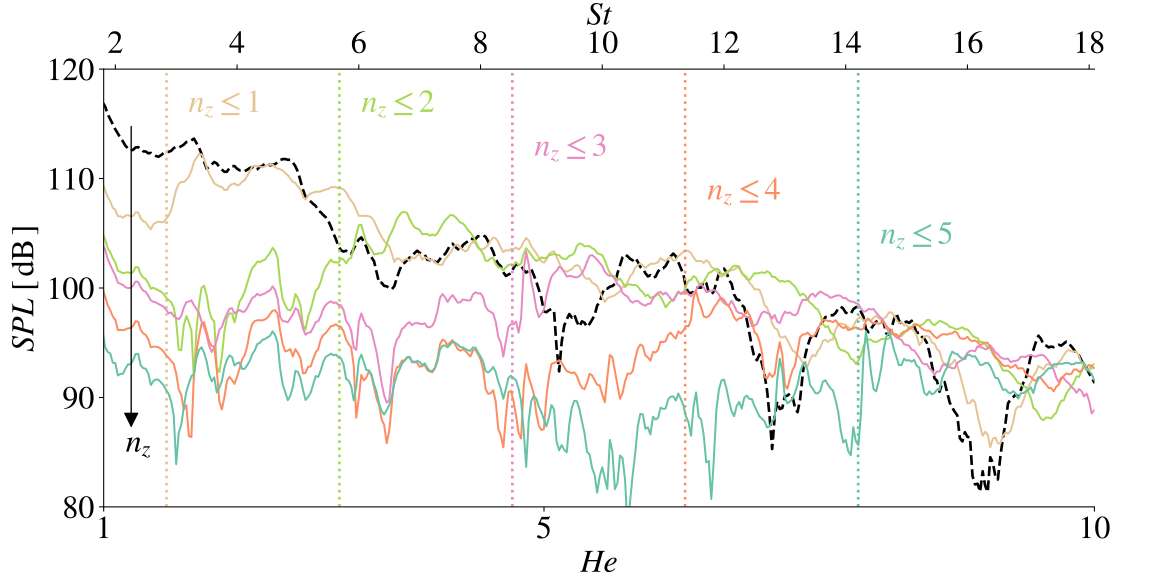}
\caption{Vertical slices of the two dimensional spectrum shown in figure~\ref{fig:scattering_condition_M5scale_c2} for the baseline case, $c2$. Each line represents a spanwise mode, from $k_z=0$ (\dashed) to $k_z=5$ (\fullline), and the vertical dotted lines indicate the discrete mode numbers allowed by the scattering condition as a function of $He$.}
\label{fig:scattering_condition_c2_slices}
\end{figure}

Figure~\ref{fig:scattering_condition_c2_slices} shows six vertical cross-sections of the two-dimensional spectrum depicted in figure~\ref{fig:scattering_condition_M5scale_c1}, corresponding to the first six spanwise modes of the baseline case. This visualisation enables a more accurate comparison of the SPLs from different modes. In agreement with previous observations, each spanwise mode of wavenumber $k_z$ displays a substantial increase in sound power for $He \geq k_z$. It is noteworthy that all sound radiating modes exhibit approximately analogous SPLs. In particular, the spanwise mode ($k_z=0$) remains a significant contributor to the radiated sound field for a large part of the frequency range associated with trailing-edge noise ($1 \leq He \leq 10$). This observation further corroborates the assumption made in previous numerical studies~\cite{Sano2019,Abreu2021} to focus on the $k_z=0$ mode.

\section{Correlation between surface and acoustic pressure fluctuations}\label{sec:correlation_SPF_acoustics}

The next part of our analysis focuses on the surface pressure fluctuations acquired by the MEMS array near the trailing edge of the airfoil, and their connection to the acoustic field investigated in the previous section. First, we examine the SPFs spectra and spanwise coherence length in \S~\ref{sec:SPFs_characteristics}, before investigating the coherence between the signals of the MEMS and line arrays for spanwise coherent fluctuations in \S~\ref{sec:corr_spf_acous}.

\subsection{SPF spectra from MEMS array}\label{sec:SPFs_characteristics}

The following section presents an investigation of the SPFs spectra measured with the MEMS sensors for all experimental cases. The primary objective of this analysis is to examine trends with respect to the driving parameters, to demonstrate the consistency of our MEMS-based measurements with literature results for NACA0012 airfoils.

\begin{figure}
    \centering
    \begin{subfigure}{0.49\textwidth}
        \centering
        \includegraphics[width=\textwidth]{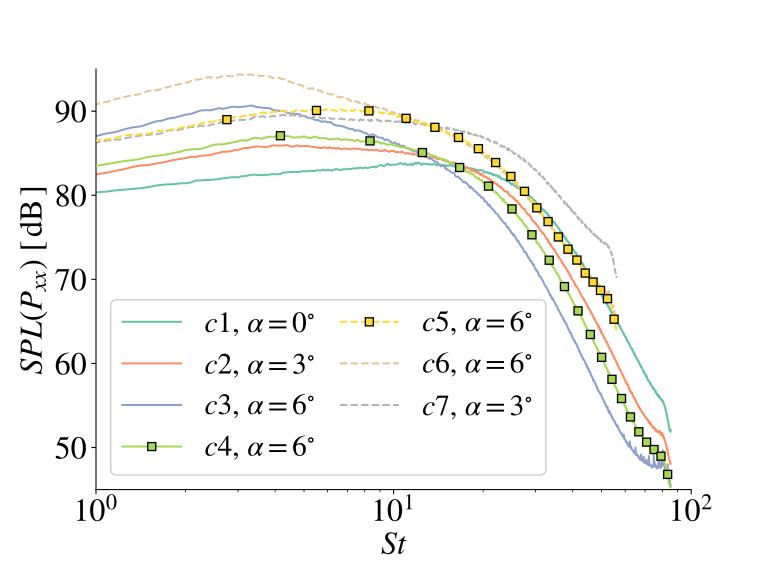}
        \caption{No scaling of $P_{xx}$.}
        \label{fig:SPF92_PSD_dim}
    \end{subfigure}
    \begin{subfigure}{0.49\textwidth}
        \centering
        \includegraphics[width=\textwidth]{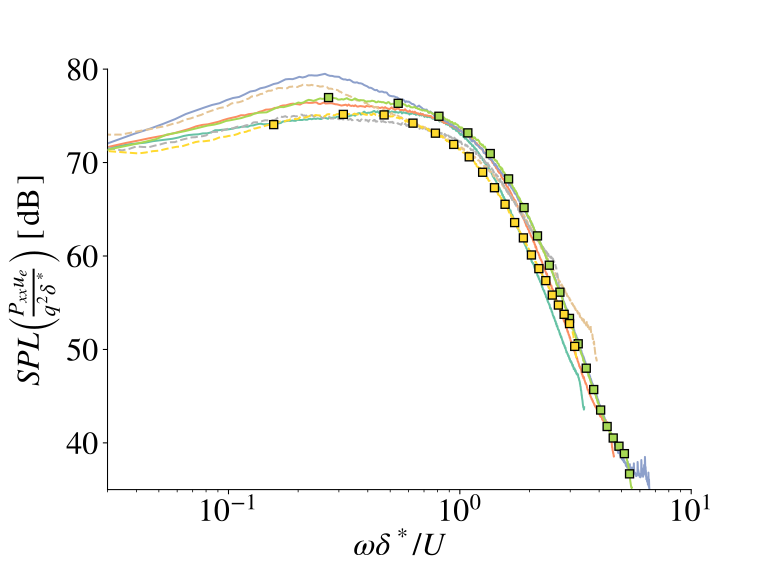} 
        \caption{Outer scaling of $P_{xx}$ and $\omega$.}
        \label{fig:SPF92_PSD_outer}
    \end{subfigure}
    \caption{SPF spectra measured from the mid-span MEMS sensor at $x/c=0.92$ for all experimental cases. Boundary-layer parameters were obtained from XFOIL~\citep{Drela1989}. Cases without tripping are highlighted by symbols (\fullsquare).}\label{fig:SPF92_PSD_scaled}
\end{figure}

Figure~\ref{fig:SPF92_PSD_scaled} shows the power spectra from the centre MEMS sensor positioned at $x/c=92\%$ on the suction side of the airfoil. The spectra from the MEMS located at $x/c=88\%$ are very similar but a few \SI{}{\decibel} lower, and thus, are not included in the figure. Since compressibility is not expected to play an important role in turbulent spectra at low Mach number~\citep{blake_mechanics_1988}, the frequency axis is given in terms of Strouhal number. Overall, the shape of the spectra and their trends with respect to changes of the angle of attack and freestream velocity are in good agreement with those reported by~\cite{herrig_broadband_2013} for a NACA0012 and by~\cite{dos_santos_wall-pressure_2023} for a NACA008 profile.
The sudden drop of the spectra at very high frequencies is non-physical and reflects the frequency response of the MEMS sensors, which is no longer flat for $f\geq$~\SI{20}{\kilo\hertz}.

The comparison of the un-tripped cases, $c4$ and $c5$, with the tripped cases at the same conditions, $c3$ and $c6$, shows an enhancement of the low-frequency contents and a more pronounced hump with tripping. This is also consistent with the parametric study conducted by~\cite{DosSantos2022}, who investigated the impact of boundary-layer tripping on the SPFs spectra. 

Figure~\ref{fig:SPF92_PSD_outer} shows the SPFs spectra scaled with the outer flow parameters, namely the boundary layer displacement thickness $\delta^*$ and edge velocity $u_e$, estimated from XFOIL at $x/c=92\%$, and the dynamic pressure $q=1/2\rho U^2$. This is in line with the work of \cite{herrig_broadband_2013} and~\cite{blake_mechanics_1988} (among others), who reported a reasonable collapse of SPFs spectra when scaling the frequency and power spectra with boundary-layer parameters obtained near the trailing edge. We find similar trends than the reported in these references, with deviations near the SPLs maxima of about \SI{5}{\decibel} for the tripped airfoil, and about \SI{2}{\decibel} for the clean one.

The preceding results demonstrate that the SPFs measured with our MEMS-based setup are in good agreement with the trends and scaling reported in literature. Note that no vibration-induced peaks were found in the spectra. Consequently, these results can be used with confidence in the following section. 

\subsection{Coherence length obtained from the MEMS array}\label{sec:coherence_length}

Next we consider the spanwise coherence of surface fluctuations near the trailing edge, determine the coherence length and compare it  with the spanwise wavelengths of the acoustic field measured with the line array. 

\begin{figure}
    \centering
    \begin{subfigure}{.48\textwidth}
        \centering
        \includegraphics[width=\textwidth]{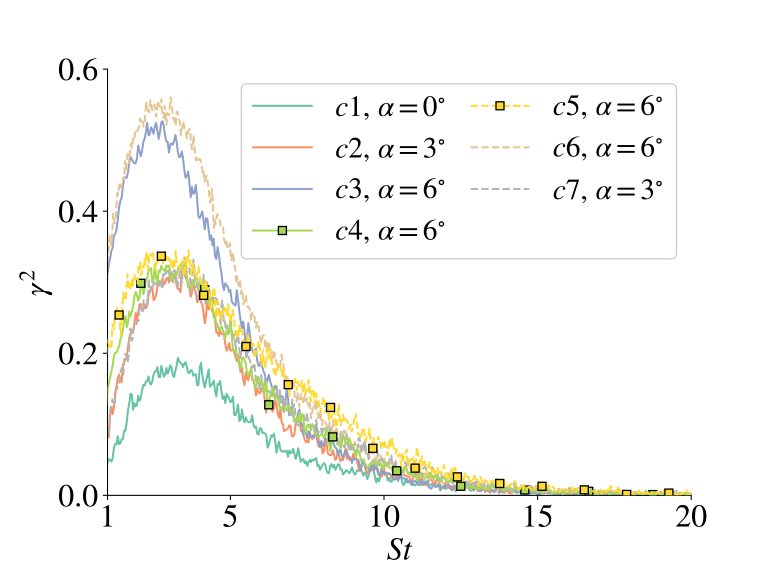}
        \caption{Coherence between span-adjacent sensors.}
        \label{fig:SpanCoherence}
    \end{subfigure}
    \begin{subfigure}{.48\textwidth}
        \centering
        \includegraphics[width=\textwidth]{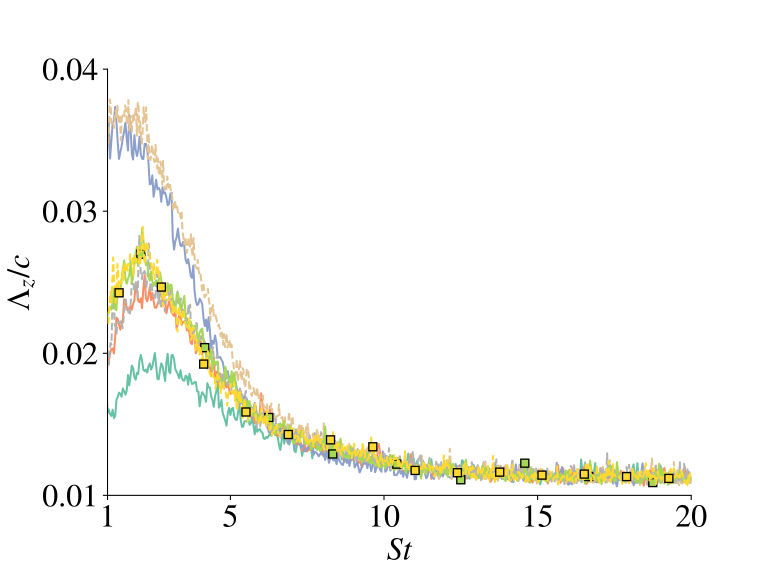} 
        \caption{Spanwise coherence length.}
        \label{fig:SpanCoherence_Length}
    \end{subfigure}
    \caption{Spanwise characteristics of the SPFs from the data collected by MEMS at $x/c=0.92$ for all experimental cases, with $Re=2\times10^5$ (\fullline) and $Re=3\times10^5$ (\dashed). Cases without tripping are highlighted by symbols (\fullsquare).}\label{fig:SpanCoherence_mems}
\end{figure}

Figure~\ref{fig:SpanCoherence} shows the coherence of the SPLs as a function of Strouhal number for all experimental cases. It is given as 
\begin{equation}\label{eq:coherence_squared}
    \gamma_{ij}^2(\omega)=\frac{|C_{ij}(\omega)|^2}{P_{ii}(\omega)P_{jj}(\omega)}\text{,}
\end{equation} 
and it is determined from two adjacent MEMS sensors separated by $\Delta z=0.02c$.

For all cases we observe the highest coherence at a similar frequency range. However, we observe strong difference of this maximum for the different cases, which correlates well with the thickness of the boundary layer at this location (see table~\ref{tab:main_parameters} for comparison). E.g. comparing tripped and non-tripped results, we observe a much lower coherence for the non-tripped case due to smaller boundary layer thickness. This is expected since a large boundary layer implies larger turbulent integral scales enhancing the coherence between two adjacent points.   

Figure~\ref{fig:SpanCoherence_Length} shows the spanwise coherence length as a function of frequency for all cases. It is given as 
\begin{equation}\label{eq:coherence_len}
    \Lambda_z(\omega)=\int_0^{\infty}\gamma_{ij}(\omega)d\Delta z\text{,}
\end{equation}
which is determined by integrating the coherence over all possible spanwise intervals between pairs of sensors $i$ and $j$, $\Delta z$, in the MEMS array located at $x/c=92\%$. The coherence length is normalized with respect to the airfoil chordlength, to provide a direct comparison with the spanwise wavenumber of the acoustic field determined from the line array, as discussed in \S~\ref{sec:acous_charac}. Considering the baseline case ($c2$), the coherence length reaches a maximum value of $\Lambda_z/c\approx2.5\%$ at $St\approx1.5$, which corresponds to $He\approx2.7$. This can be compared to the smallest spanwise wavelength of the propagative acoustic waves detected by the line array, which is $\lambda_z/c=2\pi/He=230\%$ at this frequency (see figure~\ref{fig:scattering_condition_M5scale_c2}). This clear disparity of the coherence length in the turbulent boundary layer and the spanwise wavelength of the farfield acoustics confirms the considerations of the toy problem discussed in~\S~\ref{sec:toy-model}. 
It questions the physical interpretation of the coherence length as a length scale causing TE noise and motivates our attention to flow structures with much larger spanwise coherence. 


\subsection{Correlation between acoustics and spanwise-coherent SPFs}\label{sec:corr_spf_acous}

We now examine the correlation between the SPFs measured with the MEMS and the acoustic field measured synchronously by the line array. To isolate the effect of spanwise-coherent SPFs on the acoustic field, we compare the correlations between single point SPL measurers and spanwise-averaged SPL measures (which is equal to $k_z=0$), making use of the entire MEMS array.



\begin{figure}
    \centering
    \begin{subfigure}{.49\textwidth}
        \centering
        \includegraphics[width=\textwidth]{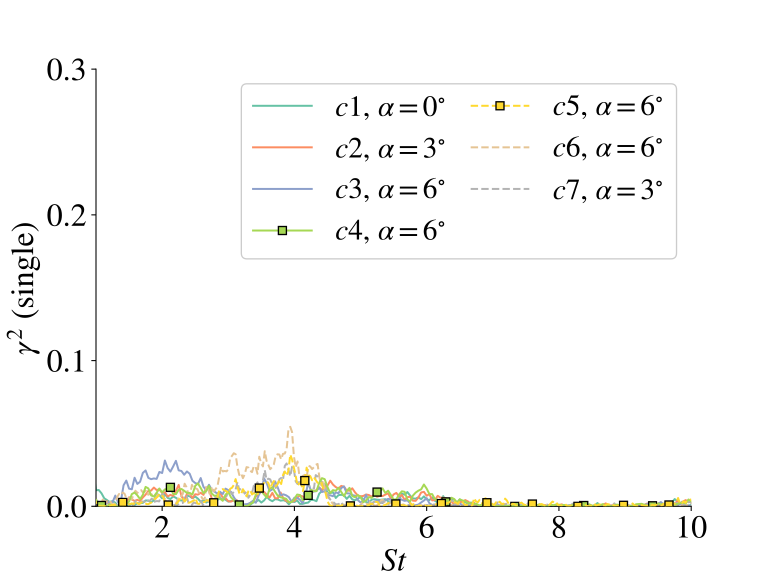}
        \caption{Pair of microphones.}
        \label{fig:coherence_MEMS1_line_single_allCases}
    \end{subfigure}
    \begin{subfigure}{.49\textwidth}
        \centering
        \includegraphics[width=\textwidth]{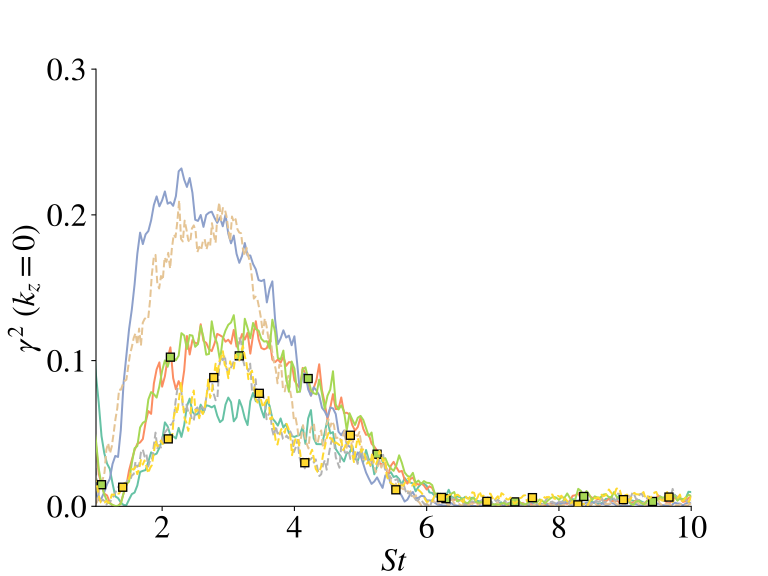} 
        \caption{Spanwise-averaged microphones ($k_z=0$).}
        \label{fig:coherence_MEMS1_line_kz0_allCases}
    \end{subfigure}
    \caption{Coherence between surface pressure fluctuations at $x/c=0.92$ and acoustic field for all experimental cases, with $Re=2\times10^5$ (\fullline) and $Re=3\times10^5$ (\dashed). Cases without tripping are highlighted by symbols (\fullsquare). Calculations considering (a) a single pair of microphones and (b) spanwise-coherent $\left(k_z=0\right)$ signals for both the surface and acoustic field measurements.}\label{fig:coherence_SPF_Line}
\end{figure}

Figure~\ref{fig:coherence_MEMS1_line_single_allCases} shows the coherence between a single MEMS sensor located at mid-span and $x/c=92\%$ with a single microphone located at the centre of the farfield line array.
For all experimental conditions, the coherence remains fairly low, providing no  evidence for a strong correlation between the SPFs and the acoustics field. Small-amplitude humps are observed for the tripped cases with $\alpha=6^{\circ}$ ($c3$ and $c6$), but coherence values remain too weak for these cases to draw conclusions. Note that applying selected band-pass filters to the time-signals provided no discernible enhancement of our results. 

In contrast, by span-averaging the MEMS signals, the correlation with the span-averaged acoustic farfield is strongly increased. As shown in figure~\ref{fig:coherence_MEMS1_line_kz0_allCases}, the coherence is significantly enhanced within $1 \leq St \leq 6$ for all cases. The maximum coherence of all cases is obtained for conditions yielding thicker boundary layers ($c3$ and $c6$), which is in line with the observation made for the spanwise coherence (see fig~\ref{fig:SpanCoherence_mems}). 

The very low coherence for the single point correlations and the dramatic enhancement for the span-averaged results clearly suggest that spanwise-coherent fluctuations causing $k_z=0$ SPFs are present in the turbulent boundary layer and are the driver for the corresponding $k_z$-component of the acoustic field.

The present results are a first experimental evidence that spanwise-coherent  $k_z=0$ modes, present in the SPFs of the fully turbulent boundary layer, are strongly correlated to the far-field acoustics for the frequency range relevant for TE noise. These findings corroborate those of the numerical study of the $k_z=0$ component of TBL-TE noise by~\cite{Sano2019}.  However, it is noteworthy that the smaller spanwise width of their numerical setup and periodic boundary conditions impose constraints that artificially boost the energy of  spanwise-coherent structures. This effect can be excluded in the present experiment. 

\subsection{Causality consideration of TE noise generation}\label{sec:time_delay}

As recalled by~\cite{jaiswal_aeroacoustic_2023}, high coherence levels between two signals does not necessary mean strong causality and must be discussed with care.
E.g. \cite{herrig_broadband_2013} noted that the back-scattering of acoustic waves from the trailing edge can artificially increase the spanwise coherence of SPFs, and could therefore be the cause for the high coherence of the  $k_z=0$ mode observed presently. To exclude this scenario and to show causality between the SPF and the far-field acoustics, we compare the time delays between signals of the $k_z=0$ SPFs and acoustic field fluctuations to theoretical time delays based on transmission scenarios (assuming causality). The goal is to ensure that our results are indeed linked to hydrodynamic coherent fluctuations radiating sound and not to back-scattering.

\begin{figure}
    \centering
    \begin{subfigure}{.45\textwidth}
        \centering
        \includegraphics[width=\textwidth]{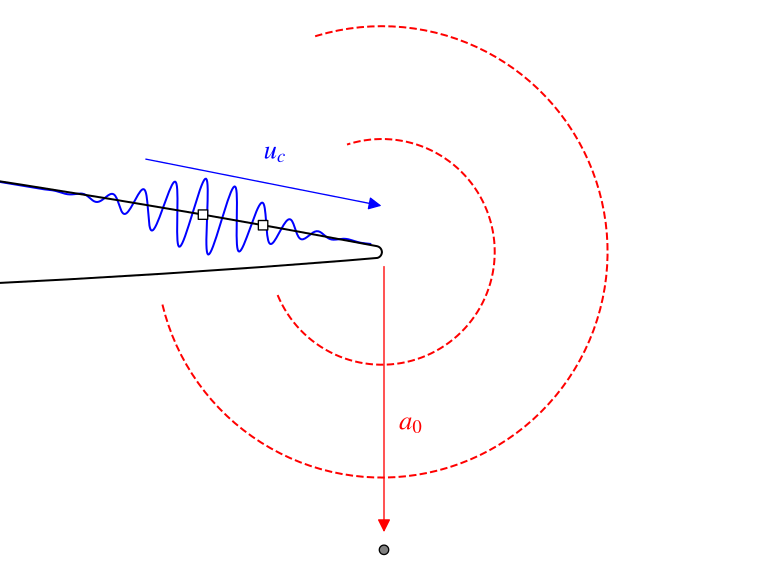}
        \caption{Convection + acoustic scattering}
        \label{fig:feedback_1}
    \end{subfigure}
    \begin{subfigure}{.45\textwidth}
        \centering
        \includegraphics[width=\textwidth]{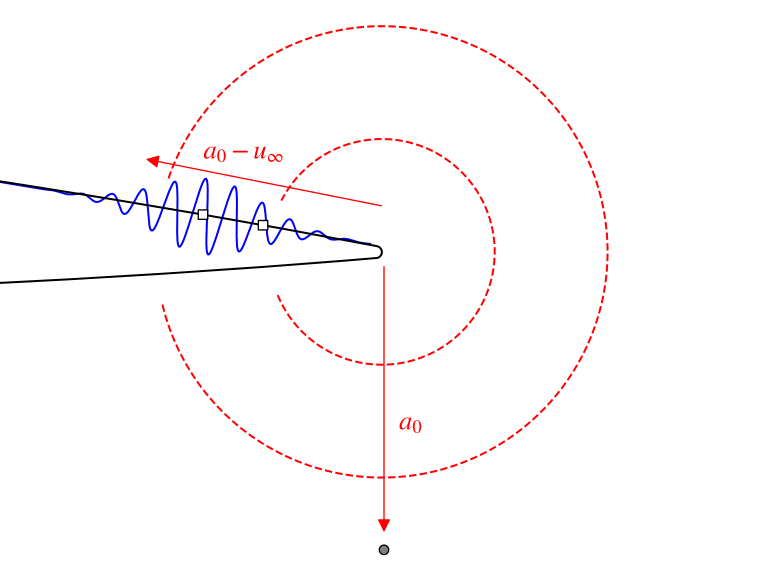} 
        \caption{Acoustic back-scattering}
        \label{fig:feedback_2}
    \end{subfigure}
    
    \begin{subfigure}{.48\textwidth}
        \centering
        \includegraphics[width=\textwidth]{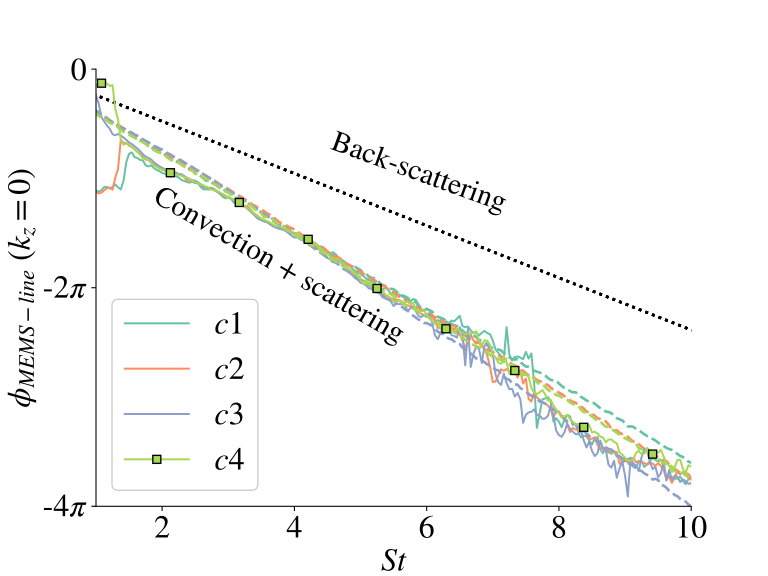}
        \caption{$Re=2\times10^5$, $M = 0.088$}
        \label{fig:time_delays_M0d088}
    \end{subfigure}
    \begin{subfigure}{.48\textwidth}
        \centering
        \includegraphics[width=\textwidth]{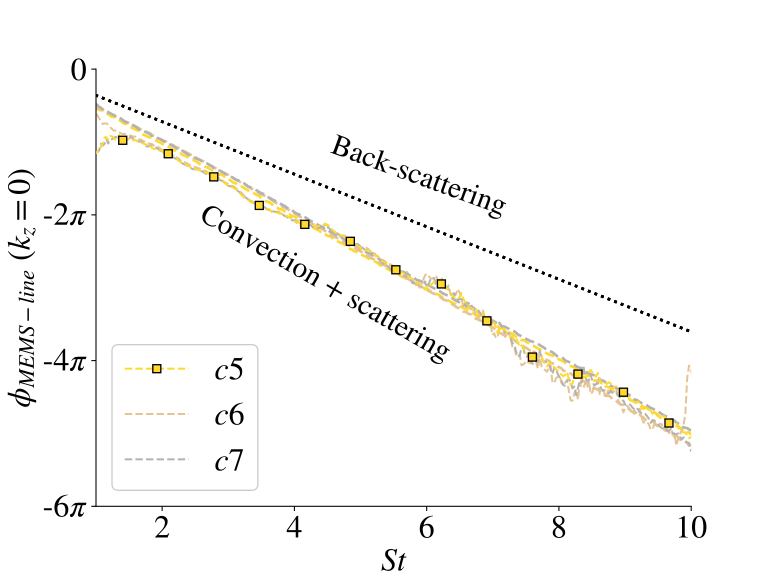} 
        \caption{$Re=3\times10^5$, $M = 0.133$}
        \label{fig:time_delays_M0d133}
    \end{subfigure}
    \caption{Time delays and phase angles of pressure waves determined from the signals of the MEMS and line array. Theoretical phase delays sketches for two scenarios: (a) $\Delta t_{cs}$~\eqref{eq:time_delay_cs}: a hydrodynamic wave is convected from the MEMS (\opensquare) to the TE where it scatters an acoustic waves towards the line array (\opencirc); (b) $\Delta t_{bs}$~\eqref{eq:time_delay_bs}: an acoustic wave generated at the TE is backscatter to the MEMS array and scattered towards the line array. Comparison with the experimental data (assuming $k_z=0$) for (c) $M=0.088$ and (d) $M=0.133$. Cases without tripping are highlighted by symbols (\fullsquare).}\label{fig:time_delays}
\end{figure}

Figures~\ref{fig:feedback_1} and~\ref{fig:feedback_2} compare the measured time delays with the theoretical time delays based on two scenarios. The first scenario considers the time delay $\Delta t_{cs}$ associated with a span-wise coherent structure being convected from the MEMS sensor to the TE and the time delay between the scattered wave propagating from the TE to the line array. The second scenario considers the time delay $\Delta t_{bs}$ of an acoustic wave being backscattered from the TE to the MEMS array and scattered from the TE to the line array. Time delays for these scenarios are obtained as
\begin{subequations}\label{eq:time_delays}
    \begin{equation}\label{eq:time_delay_cs}
        \Delta t_{cs} =d_M/u_c(\omega) + d_L/a_s(\omega)\text{,}
    \end{equation}
    \begin{equation}\label{eq:time_delay_bs}
        \Delta t_{bs} = d_L/a_s(\omega) - d_M/(a_s(\omega)-U_e)\text{,}
    \end{equation}
\end{subequations}
where $d_M=0.08c$ is the distance from the MEMS line to the TE, $d_L=3c$ is the vertical distance between the TE and the acoustic line array, $U_e$ is the mean flow velocity at the boundary layer edge (obtained from XFOIL for each case), and $u_c(\omega)$ is the convection velocity associated with SPFs, which is a function of the frequency. The latter is obtained from the phase difference between the two available MEMS arrays, as shown in appendix~\ref{sec:SPFs_stream_characteristics}. The comparison is shown for the two Mach numbers $M=0.088$ and $M=0.133$. The measured time delays are in very good agreement with the first scenario, which lends strong credibility to the conclusion that the coherence between the SPFs and the acoustic field is not the result of backscattering. 

\section{Conclusions}\label{sec:conclusions}
This study presents an experimental investigation of turbulent boundary-layer trailing-edge (TBL-TE) noise for a NACA0012 airfoil. The boundary layer was tripped to turbulence on both sides of the airfoil for certain conditions to avoid tonal noise. The chord-based Reynolds numbers were within $Re\in[2,\,3]\times10^5$. This setup produces broadband TE noise comparable to that dominating the acoustic signature typically observed in wind turbines~\citep{oerlemans_aeroacoustic_2004}. To study this phenomenon, synchronous measurements of surface and acoustic pressure fluctuations were conducted using custom spanwise microphone arrays located in the far-field and MEMS sensors positioned at the surface near the airfoil’s trailing edge.

One of the key contributions of this work is the spanwise modal decomposition of the acoustic field associated with TBL-TE noise and its correlation with span-averaged surface pressure fluctuations. This analysis reveals that coherent flow structures with large spanwise wavelengths are the primary driver of broadband noise. For low-to-mid frequencies ($He\in[1,\,10]$), which includes the peak of SPLs associated with TE noise, the radiated acoustic waves exhibit spanwise wavelengths of at least $60\%$ of the airfoil chord length. This finding aligns closely with the theoretical TE scattering condition proposed by~\cite{Nogueira2017} and alluded to by~\cite{amiet_acoustic_1975}, which predicts that only spanwise wavenumbers below the acoustic wavenumber contribute significantly to the acoustic spectrum.

This work provides a new perspective on the length scales of flow structures driving TBL-TE noise, which differ significantly from the classical approach based on the spanwise coherence length ($\Lambda_z$) of the turbulent spectrum~\citep{lee_turbulent_2021}. For the presently studied conditions, $\Lambda_z$ is about $1\%$ of the chord length, a value comparable to the boundary-layer thickness. However, $\Lambda_z$ is a statistical measure reflecting the most energetic (most probable) turbulent scale, whether or not it participates in sound generation. By contrast, the radiating structures identified here are orders of magnitude larger with respect to the spanwise length. The toy model presented in appendix~\ref{sec:SPFs_stream_characteristics} further illustrates how the superposition of large-scale structures can yield smaller coherence lengths, emphasizing that $\Lambda_z$ should not be interpreted as the characteristic length scale of the noise-driving structures.


The connection between surface and acoustic pressure fluctuations was further explored by isolating the spanwise-coherent modes. We observed a significantly increased coherence between the span-averaged surface and far-field acoustic signals compared to single point measurements, confirming that spanwise-coherent fluctuations generate a substantial portion of TBL-TE noise. Time-delay measurements validated the causality of these results, by matching theoretical predictions for waves scattered at the trailing edge, ensuring that the observed coherence reflected genuine acoustic generation rather than backscattered acoustics.

While the spanwise modal decomposition method offers valuable insights, it also presents challenges. Resolving low-wavenumber structures requires large numerical domains in simulations and high-aspect-ratio wings with extensive sensor arrays in experiments, both of which are resource-intensive. However, the method aligns well with linear mean-field analyses, such as the resolvent framework, which can extract coherent fluctuations of arbitrary spanwise wavenumber from spanwise-averaged turbulent mean flows without the need for large computational domains. These linear analyses provide physics-based models that can enhance understanding of the mechanisms driving TE noise~\citep{demange_wavepackets_2024} and support the development of effective control strategies.

The findings from this experimental study are complemented by a companion numerical analysis \citep{Yuan2024arxiv} of the same setup. That study demonstrates that the coherent structures driving TBL-TE noise manifest as streamwise-travelling wavepackets within the turbulent boundary layer, further validating the present experimental observations. Together, these works offer new insights into the mechanisms of TBL-TE noise, with implications for improving noise prediction models and developing effective noise mitigation strategies.

\section*{Acknowledgments}
Simon Demange would like to acknowledge Deutsche Forschungsgemeinschaft for supporting current work under grant number 458062719. Zhenyang Yuan would like to acknowledge Swedish Research Council for supporting current work under Grant 2020-04084.

\appendix

\section{Toy problem for coherence length of synthetic signals}\label{sec:coherence_synthetic}

We devise a simple toy-model to illustrate the connection and difference between the spanwise coherence length and the spanwise wavelength of structures composing surface pressure fluctuations in a turbulent boundary layer. In particular, we aim to show that the superposition of signals with \textit{large} wavelengths can still result in \textit{small} coherence length.

\begin{figure}
\centering
\begin{subfigure}{1\textwidth}
    \centering
    \includegraphics[width=\columnwidth]{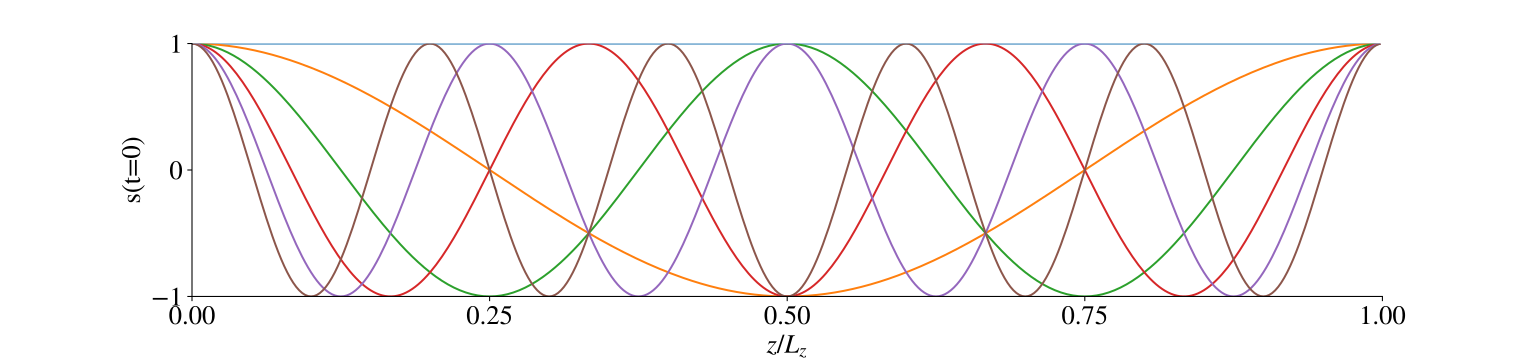}
    \caption{Mode shapes for $N=6$.}
    \label{fig:sprinkler_mode_shapes}
\end{subfigure}

\begin{subfigure}{.49\textwidth}
    \centering
    \includegraphics[width=\columnwidth]{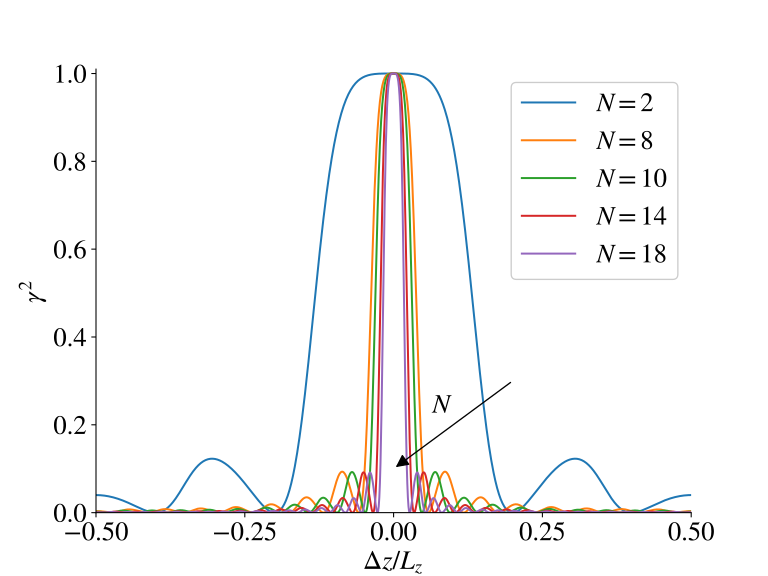} 
    \caption{Coherence.}
    \label{fig:sprinkler_span_coherence}
\end{subfigure}
\begin{subfigure}{.49\textwidth}
    \centering
    \includegraphics[width=\columnwidth]{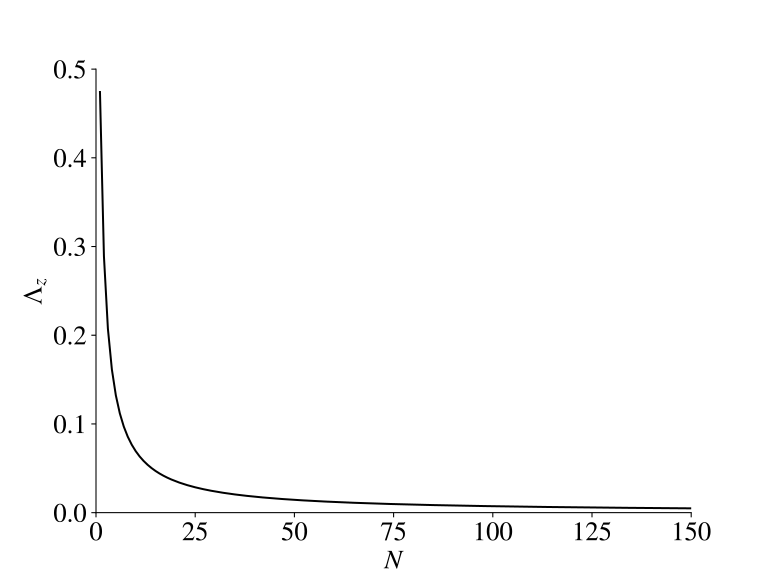} 
    \caption{Coherence length.}
    \label{fig:sprinkler_span_corLength}
\end{subfigure}
\caption{Toy model for the illustration of coherence length versus length scale of structures captured by an array of sensors: (a) mode shapes along $z$ at $t=0$ considered when $N=6$ (only positive mode numbers $n_{\text{z}}\geq0$ are shown); (b) spanwise coherence between sensors separated by $\Delta z$, for an increasing number of Fourier modes in the signals; (c) Corresponding spanwise coherence length averaged over temporal frequency as a function of the number of modes.}\label{fig:sprinkler}
\end{figure}

We consider here the synthetic signals measured by an array of sensors distributed along a line of length $L_{\text{z}}$, described by the coordinate $z\in[-L_{\text{z}}/2,\,L_{\text{z}}/2]$. For simplicity, these signals are defined as a superposition of \emph{cosine} functions with unit amplitude along the spanwise direction, each multiplied by a white-noise time signal, $s_{n_z}(t)$,
\begin{equation}\label{eq:model_signal2}
	s(z, t)=\sum_{n_{z}=-N}^{N} s_{n_z}(t)\cos(2\pi n_z z),
\end{equation}
\noindent where $k_{\text{z}}=2\pi n_{z}$ is the wavenumber along $z$ of the $n_{z}$-th mode with wavelength $\lambda_{\text{z}}=2\pi/k_{\text{z}}$ and $N$ is the total number of \emph{spanwise modes} composing the signal. The shapes of the first six modes at a given time are shown in figure~\ref{fig:sprinkler_mode_shapes} for illustration. We further assume that the time-signals from each spanwise position are uncorrelated, such that in the frequency domain $E\{s_i(\omega),~s_j(\omega)\}=\delta_{ij}$, where $E\{\cdot,\,\cdot\}$ denotes the expectancy operator and $\delta_{ij}$ is the Kronecker delta for different spanwise positions. 



Similarly to the post-treatment of experimental or numerical signals, we compute the coherence, $\gamma$, of signals from a pair of sensors separated by $\Delta z$ along the line at a given angular frequency, $\omega$, as the cross-spectrum between these signals, $P_{xy}$, normalised by the product of their auto spectra, $P_{xx}$ and $P_{yy}$,
\begin{equation}\label{eq:coherence2}
	\gamma_{xy}^2(\omega)=\frac{|P_{xy}(\omega)|^2}{P_{xx}(\omega)P_{yy}(\omega)}.
\end{equation}

For the analytical signals considered here, an analytical solution for $\gamma_{xy}^2$ with respect to the centre sensor at $z=0$ is readily obtained as
\begin{equation}\label{eq:coherence3}
	\gamma_{xy}^2=\left(\cos(N \xi) + \frac{\sin(N \xi)}{\tan(\xi/2)}\right)^2(2N+1)^{-1}\left(\frac{\sin(2N\xi + \xi)}{2\sin(\xi)}+N+\frac{1}{2}\right)^{-1},
\end{equation}
 \noindent with $\xi = 2\pi \Delta z$.

Figure~\ref{fig:sprinkler_span_coherence} shows the resulting squared coherence as a function of sensor distances $\Delta z$, for increasing total numbers of modes $N$. The coherence length, $\Lambda_{\text{z}}$, is then obtained by integrating $\gamma$ along the $\Delta z$-axis and is shown in figure~\ref{fig:sprinkler_span_corLength}. 

This model clearly shows the trivial result that the more modes with decreasing wavelengths are summed together to generate the signal, the shorter the coherence length becomes. Nonetheless, large structures are still part of the signal, with wavelengths significantly larger than the coherence length. For example, when $N=6$, which corresponds to the modes shown in Figure~\ref{fig:sprinkler_mode_shapes}, the coherence length is $\gamma^2\approx0.1$. 

Coherence decay with increasing spanwise spacing $\Delta z$ can be understood as various modes along the z direction are simultaneously present in the data; as such modes have the same amplitude, there is not a dominant wavenumber and thus one cannot obtain a deterministic phase (and thus a significant coherence) between separated probes.

\section{Streamwise correlation of SPFs}\label{sec:SPFs_stream_characteristics}

This section reports the results of the correlation between the signals of MEMS sensors from each of the two spanwise lines, located at $x/c=88\%$ and $x/c=92\%$ respectively. These results are used to obtain the streamwise convection velocity of SPFs near the trailing edge, used in section~\ref{sec:corr_spf_acous} to compute the time delays between MEMS and line array signals.

\begin{figure}
    \centering
    \begin{subfigure}{.48\textwidth}
        \centering
        \includegraphics[width=\textwidth]{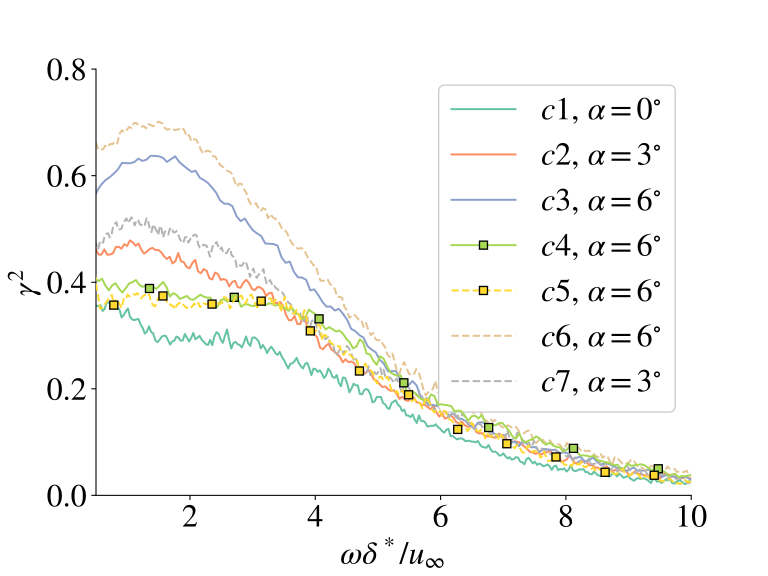}
        \caption{Coherence}
        \label{fig:streamwise_coherence_SPFs}
    \end{subfigure}
    \begin{subfigure}{.48\textwidth}
        \centering
        \includegraphics[width=\textwidth]{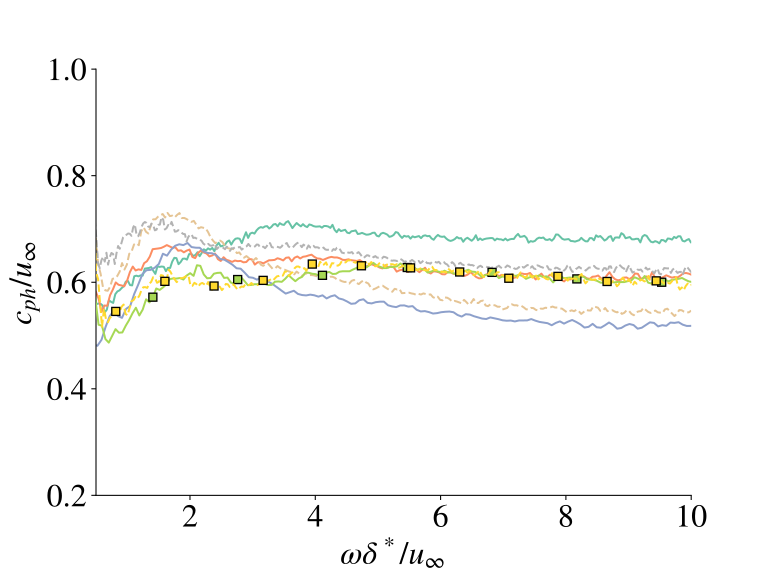} 
        \caption{Phase speed}
        \label{fig:streamwise_phase_speed_SPFs}
    \end{subfigure}
    \caption{Streamwise characteristics of the spanwise-coherent ($k_z=0$) SPFs from the data collected by MEMS at $x/c=0.88$ and $x/c=0.92$ for all experimental cases. (a) Coherence between a single streamwise pair of sensors ($\dashed$) and between span-averaged signals ($\fullline$); (b) Resulting streamwise phase speed normalised by the free stream velocity for span-averaged signals.}\label{fig:streamwise_SPFs}
\end{figure}

Figure~\ref{fig:streamwise_coherence_SPFs} illustrates the coherence between the spanwise-averaged signals ($k_z=0$) of each line of MEMS sensors for all experimental conditions. As with the spanwise coherence (see figure~\ref{fig:SpanCoherence}), an increase in the boundary-layer thickness caused by either higher angles of attack or tripping results in an increase in coherence. 

The streamwise convection velocity of structures in the turbulent boundary layer is calculated from~\eqref{eq:cph_x}, and shown in figure~\ref{fig:streamwise_phase_speed_SPFs} as a function of frequency. The resulting phase velocity distributions are similar to that presented by~\citet{brooks_trailing_1981} for a NACA0012 airfoil at a higher Reynolds numbers, and are found to be mostly around $c_{ph}/u_{\infty}=60\%$ for the frequencies investigated. Note that very similar results were obtained when considering the signals from pairs of sensors, rather than the span-averaged signals, which seemingly indicates that $k_z=0$ structures are generally travelling at similar convection velocities than the bulk of turbulent structures. 



\bibliographystyle{jfm}
\bibliography{references}

\end{document}